\documentclass[runningheads,a4paper]{llncs}

\usepackage{graphicx}
\usepackage[]{xparse}
\usepackage[]{xspace}
\usepackage{amsthm, amsmath, amssymb}
\usepackage{mathtools}
\usepackage{stmaryrd}
\usepackage{xifthen}
\usepackage{varwidth}
\usepackage{subfiles}
\usepackage[]{subcaption}
\usepackage[T1]{fontenc}
\usepackage[utf8]{inputenc}
\usepackage{lmodern}
\usepackage{calc}
\usepackage{xifthen}
\usepackage{booktabs}

\usepackage{xcolor}

\definecolor{lightgray}{rgb}{0.95,0.95,0.95}
\definecolor{darkgray}{gray}{0.45}

\definecolor{highlight1}{HTML}{50d890}
\definecolor{highlight2}{HTML}{4f98ca}

\usepackage{tikz}
\newlength\vgap
\newlength\hgap
\newlength\helper

\newcommand{\hsp}{\vphantom{Ag}}

\usetikzlibrary{arrows,positioning,fit,shapes,backgrounds,patterns,shapes.misc}
\tikzset{%
  object/.style={
    draw=black,
    inner sep=0pt,
    minimum size=5pt,
    shape=circle,
    fill=black
  },
  rolein/.style={
    draw=black,
    arrows={stealth-},
    thick,
    shorten <=2pt,
    shorten >=2pt
  },
  roleout/.style={
    draw=black,
    arrows={-stealth},
    thick,
    shorten <=2pt,
    shorten >=2pt
  },
  compact/.style={
	  inner sep=0.4ex,
	  font=\sffamily\scriptsize\hsp
	},
  iri/.style={
    draw=darkgray,
    fill=white,
    rectangle,
    rounded corners,
    thick,
    text centered,
    minimum width=0.5cm,
    font=\sffamily\small\hsp
	},
  var/.style={
    draw=darkgray,
    fill=white,
    rectangle,
    rounded corners,
    thick,
    dotted,
    text centered,
    minimum width=0.5cm,
    font=\sffamily\small\hsp
	},
  smalliri/.style={
    draw=darkgray,
    fill=white,
    rectangle,
    rounded corners,
    thick,
    text centered,
    minimum width=0.4cm,
    font=\sffamily\tiny\hsp
	},
  literal/.style={
    draw=darkgray,
    fill=white,
    rectangle,
    thick,
    text centered,
    font=\sffamily\small\hsp
	},
  label/.style={
     text centered,
     anchor=center,
     fill=white,
     opacity=0.98,
     text opacity=1,
     inner sep=0.2ex,
     font=\sffamily\footnotesize\hsp
	},
  arrin/.style={
    draw=darkgray,
    arrows={stealth-},
    thick,
    font=\sffamily\footnotesize\hsp,
    pos=0.55
	},
  arrout/.style={
    draw=darkgray,
    arrows={-stealth},
    thick,
    font=\sffamily\footnotesize\hsp,
	  pos=0.45
	},
  block/.style={
    rectangle,
    dashed,
    rounded corners,
    draw=darkgray,
    thick,
    pattern=crosshatch,
    pattern color=lightgray,
    inner sep=7pt
  },
  block1/.style={
    rectangle,
    dashed,
    rounded corners,
    draw=darkgray,
    thick,
    pattern=north west lines,
    pattern color=lightgray,
    inner sep=7pt
  },
  block2/.style={
    rectangle,
    dashed,
    rounded corners,
    draw=darkgray,
    thick,
    pattern=north east lines,
    pattern color=lightgray,
    inner sep=7pt
  },
  oval/.style={
    rectangle,
    dashed,
    rounded corners,
    draw=darkgray,
    thick,
    inner sep=25pt
  },
  component/.style={
    align=center,
    draw=darkgray,
    fill=white,
    rectangle,
    rounded corners,
    thick,
    text centered,
    dashed,
    minimum width=0.5cm,
    font=\sffamily\scriptsize\hsp
	},
  built/.style={
    draw=darkgray,
    dashed,
    arrows={latex-},
    thick,
    shorten <=2pt,
    shorten >=2pt
	},
}

\usepackage[]{listings}

\newlength\listingnumberwidth
\newlength{\llin}
\lstdefinestyle{lst}{
	backgroundcolor=\color{white},
  basicstyle={\ttfamily},
  keywordstyle=\color{darkgray},
  commentstyle=\color{darkgray}\ttfamily,
  tabsize=2,
  numbers=left,
  numberstyle=\color{darkgray},
  showtabs=false,
  showspaces=false,
  showstringspaces=false,
  extendedchars=true,
  breaklines=true,
	columns=fixed, 
	basewidth=0.5em, 
  xleftmargin=\listingnumberwidth,
	moredelim=[is][\bfseries]{~}{~},
  mathescape=true,
  escapechar=`
}

\lstdefinelanguage{rdfn3}{
	keywords={rdf,sh,rdfs,ex,owl},
  morestring=[b]",
}

\lstdefinelanguage{dl}{
  morecomment=[l]{//}, 
  mathescape=true,
}

\newlength\wordWidth

\NewDocumentCommand{\tikzedgevarwidth}{ O{arrin} O{iri} O{iri} m m m }{%
  \setlength{\wordWidth}{\widthof{{\scriptsize #5}}+0.5cm}%
  \begin{tikzpicture}[baseline=-3pt]\node[#2,compact](A){#4};%
    \node[#3,compact,right=\wordWidth of A]{#6} edge[#1] node[label,xshift=-1pt] {#5} (A);%
  \end{tikzpicture}%
}

\NewDocumentCommand{\tikzedgenonodes}{O{1.55cm} O{white} O{arrin} m }{%
  \setlength{\wordWidth}{\widthof{{\scriptsize #4}} + 0.5cm}%
  \begin{tikzpicture}[baseline=-3pt]%
    \node[inner sep=0](A){};%
    \node[inner sep=0,right=\wordWidth of A]{} edge[#3] node[label,fill=#2,xshift=-1pt] {\scriptsize #4} (A);%
  \end{tikzpicture}%
}

\newcommand{\tikznode}[2]{\begin{tikzpicture}[baseline=-3pt]\node[#1,inner sep=0.4ex]{\scriptsize #2};\end{tikzpicture}}

\NewDocumentCommand{\gedgenonodes}{ O{1.55cm} O{white} O{arrin} m }{\tikzedgenonodes[#1][#2][#3]{#4}}
\NewDocumentCommand{\gedge}{ O{iri} O{iri} m m m}{\tikzedgevarwidth[arrin][#1][#2]{#3}{#4}{#5}}
\NewDocumentCommand{\gnode}{ O{iri} m}{\tikznode{#1}{#2}}

\NewDocumentCommand{\term}{m}{\texttt{#1}}

\NewDocumentCommand{\shapename}{m}{\term{#1}}

\newcommand{\constrToConceptExpr}{\ensuremath \tau_\mathrm{constr}\xspace}
\newcommand{\targetNoToConceptExpr}{\ensuremath \tau_\mathrm{target}\xspace}
\newcommand{\shapesToKB}{\ensuremath \tau_\mathrm{shapes}\xspace}
\newcommand{\pathToRole}{\ensuremath \tau_\mathrm{role}\xspace}
\newcommand{\contained}[2]{\ensuremath #1~{<:}_{S}~#2\xspace}
\newcommand{\pe}{\ensuremath\rho\xspace}

\newcommand{\constructed}[2]{\ensuremath #1^{\text{\guilsinglleft}#2\text{\guilsinglright}}\xspace}

\newcommand{\toAtomicConcept}[1][]{%
  \ifthenelse{\isempty{#1}}%
  {\ensuremath \tau_\mathrm{name}\xspace}%
  {\ensuremath \tau_\mathrm{name}(#1)\xspace}%
}
\newcommand{\toObjectName}[1][]{%
  \toAtomicConcept[#1]
}
\newcommand{\toGraphNode}[1][]{%
  \ifthenelse{\isempty{#1}}%
  {\ensuremath \tau_\mathrm{node}\xspace}%
  {\ensuremath \tau_\mathrm{node}(#1)\xspace}%
}

\newcommand{\mo}[1]{\ensuremath\mathrm{Mod}(#1)}
\newcommand{\finmo}[1]{\ensuremath\mathrm{Mod}^\mathrm{fin}(#1)}
\newcommand{\names}[1]{%
  \ifthenelse{\isempty{#1}}%
  {\ensuremath\mathrm{Names}\xspace}%
  {\ensuremath\mathrm{Names}(#1)}%
}%
\newcommand{\faith}[1]{\ensuremath\mathrm{Faith}(#1)}

\NewDocumentCommand{\evalq}{m O{G}}{\ensuremath\llbracket #1 \rrbracket_{#2}}

\NewDocumentCommand{\evalc}{O{v} O{G} O{\sigma} m}%
{\ensuremath\llbracket #4 \rrbracket^{#1, #2, #3}}%

\NewDocumentCommand{\evalp}{m O{G}}{\ensuremath\llbracket #1 \rrbracket^{#2}}%

\NewDocumentCommand{\restr}{}{{\text{restr}}}%
\NewDocumentCommand{\ninv}{}{{\text{non-inv}}}%

\newif\ifextended

\newcommand{\ifExtended}[2]{%
  \ifextended%
  #1%
  \else%
  #2%
  \fi
}%

\begin{document}
 
\extendedtrue

\ifExtended{
  \title{Deciding SHACL Shape Containment through Description Logics Reasoning \\ (Extended Version)}
}{
  \title{Deciding SHACL Shape Containment through Description Logics Reasoning}
}
\titlerunning{Deciding SHACL Shape Containment through DL Reasoning}

\author{
  Martin Leinberger\inst{1}\and
  Philipp Seifer\inst{2} \and
  Tjitze Rienstra\inst{1} \and
  Ralf Lämmel\inst{2}\and
  Steffen Staab\inst{3,4}
}
\authorrunning{M. Leinberger et al.}
\institute{
  Inst. for Web Science and Technologies, University of Koblenz-Landau, Germany \and
  The Software Languages Team, University of Koblenz-Landau, Germany \and
  Institute for Parallel and Distributed Systems, University of Stuttgart,
  Germany \and
  Web and Internet Science Research Group, University of Southampton, England
}

\maketitle
\begin{abstract}
  The Shapes Constraint Language (SHACL) allows for formalizing constraints
  over RDF data graphs. A shape groups a set of constraints that may be fulfilled
  by nodes in the RDF graph. We investigate the problem of containment
  between SHACL shapes. One shape is contained in a second shape if every graph node
  meeting the constraints of the first shape also meets the constraints of the second. To
  decide shape containment, we map SHACL shape graphs into description logic
  axioms such that shape containment can be answered by description logic
  reasoning. We identify several, increasingly tight syntactic restrictions of
  SHACL for which this approach becomes sound and complete.
\end{abstract}

\section{Introduction}
\label{sec:introduction}

RDF has been designed as a flexible, semi-structured data format. To ensure data
quality and to allow for restricting its large flexibility in specific domains,
the W3C has standardized the \emph{Shapes Constraint Language
(SHACL)}\footnote{https://www.w3.org/TR/shacl/}. A set of SHACL shapes are
represented in a \emph{shape graph}. A shape graph represents constraints that
only a subset of all possible RDF data graphs \emph{conform} to. A SHACL processor
may validate whether a given RDF data graph conforms to a given SHACL
shape graph.

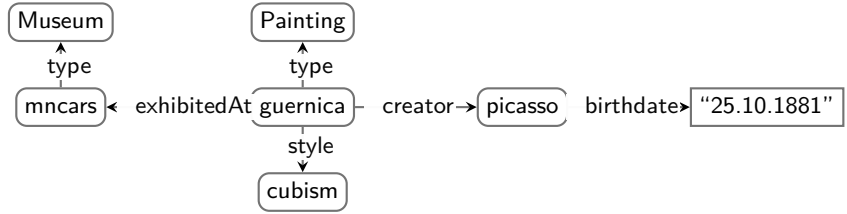
\begin{figure}[tb]
  \begin{subfigure}{1.0\textwidth}
\begin{lstlisting}[style=lst, language=rdfn3]
:PaintingShape a sh:NodeShape;
  sh:targetClass :Painting;
  sh:property [ sh:path :exhibitedAt; sh:minCount 1; ];
  sh:property [ sh:path :creator; sh:node :PainterShape; ].

:PainterShape a sh:NodeShape;
  sh:property [ sh:inversePath :creator; sh:node :PaintingShape; ];
  sh:property [ sh:path :birthdate; sh:minCount 1; sh:maxCount 1; ];

:CubistShape a sh:NodeShape
  sh:property [ sh:path ( [sh:inversePath :creator] :style );
      sh:minCount 1; sh:value :cubism; ].
\end{lstlisting}
    \subcaption{Example for a SHACL shape graph.}
    \label{subfig:shape_graph_intro}
  \end{subfigure}

  \vspace{0.25cm}

  \begin{subfigure}{1.0\textwidth}
    \setlength{\vgap}{0.6cm}
    \setlength{\hgap}{1.6cm}
    \centering
    \begin{tikzpicture}
      \node[iri] (Painting) {Painting};
      \node[iri, left=1.8cm of Painting] (Museum) {Museum};

      \node[iri, below=\vgap of Painting] (guernica) {guernica}
      edge[arrout] node[label]{type} (Painting);

      \node[iri, right=\hgap of guernica] (picasso) {picasso}
      edge[arrin] node[label]{creator} (guernica);


      \node[literal, right=\hgap of picasso] {``25.10.1881''}
      edge[arrin] node[label]{birthdate} (picasso);

      \node[iri, below=\vgap of Museum] (mncars) {mncars}
      edge[arrin] node[label]{exhibitedAt} (guernica)
      edge[arrout] node[label]{type} (Museum);

      \node[iri, below=\vgap of guernica] (cubism) {cubism}
      edge[arrin] node[label]{style} (guernica);



    \end{tikzpicture}
    \subcaption{Example for a data graph that conforms to the shape graph.}
    \label{subfig:data_graph_intro}
  \end{subfigure}
%
%
%
  \caption{Example of a shape graph (a) and a data graph (b).}
  \label{fig:example_validation_intro}
\end{figure}

A shape graph and a data graph that act as a running example are presented in
Fig.~\ref{fig:example_validation_intro}. The shape graph introduces
a \shapename{PaintingShape} (line 1--4) which constrains all instances of the
class \gnode{Painting}. It requires the presence of at least one
\gedgenonodes{exhibitedAt}~property (line 3) as well as that each node reachable
via the \gedgenonodes{creator}~property from a \gnode{Painting} conforms to the
\shapename{PainterShape} (line 4). The \shapename{PainterShape} (lines 5--8)
requires all incoming \gedgenonodes{creator}~properties to conform to
\shapename{PaintingShape} (line 6) as well as the presence of exactly one
\gedgenonodes{birthdate}~property. Lastly, the shapes define
a \term{CubistShape} (lines 9--11) which must have an incoming
\gedgenonodes{creator}~property from a node that has an outgoing
\gedgenonodes{style}~property to the node \gnode{cubism}. The graph shown in
Fig.~\ref{fig:example_validation_intro} conforms to this set of shapes as it
satisfies the constraints imposed by the shape graph.

In this paper, we investigate the problem of \emph{containment} between shapes:
Given a shape graph $S$ including the two shapes $s$ and $s'$, intuitively $s$
is contained in $s'$ if and only if every data graph node that conforms to $s$
is also a node that conforms to $s'$. An example of a containment problem is the
question whether \shapename{CubistShape} is contained in
\shapename{PainterShape} for all possible RDF graphs. While containment is not
directly used in the validation of RDF graphs with SHACL, it offers means to
tackle a broad range of other problems such as SHACL constraint debugging, query
optimization~\cite{semanticQueryOptimization,optimisingWithShexConstraints,shexQueryContainment} 
or program verification~\cite{tycus}. As an example of query optimization,
assume that \shapename{CubistShape} is contained in \shapename{PainterShape} and
that the graph being queried conforms to the shapes.  A query querying for
\gnode[var]{?X} and \gnode[var]{?Y} such that
\gedge[var][iri]{?X}{style}{cubism}, \gedge[var][var]{?X}{creator}{?Y} and
\gedge[var][var]{?X}{exhibitedAt}{?Z} can be optimized. Since nodes that are
results for \gnode[var]{?Y} must conform to \shapename{CubistShape} and
\shapename{CubistShape} is contained in \shapename{PainterShape}, nodes that are
results for \gnode[var]{?X} must conform to \shapename{PaintingShape}.
Subsequently, the pattern \gedge[var][var]{?X}{exhibitedAt}{?Z} can be removed
without consequence. 

Given a set of shapes $S$, checking whether a shape $s$ is
contained in another shape $s'$ involves checking whether there is no
counterexample. That means, searching for a graph that conforms to $S$, but in
which a node exists that conforms to $s'$ but not to $s$.
A similar problem is concept subsumption in description logics (DL). For DL,
efficient tableau-based approaches~\cite{introdl} are known that either disprove
concept subsumption by constructing a counterexample or prove that no
counterexample can exist. Despite the fundamental differences between the
Datalog-inspired semantics of SHACL~\cite{shaclSemantics} and the Tarski-style
semantics used by description logics, we leverage concept subsumption in
description logic by translating SHACL shapes into description logic knowledge
bases such that the shape containment problem can be answered by performing
a subsumption check.

\paragraph*{Contributions} We propose a translation of the
containment problem for SHACL shapes into a DL concept subsumption problem such
that the formal semantics of SHACL shapes as defined in~\cite{shaclSemantics} is
preserved. Our contributions are as follows:
\begin{enumerate}
  \item We define a syntactic translation of a set of SHACL shapes into a description
    logic knowledge base and show that models of this knowledge base and the
    idea of faithful assignments for RDF graphs in SHACL can also be mapped into
    each other.
  \item We show that by using the translation, the containment of SHACL shapes can
    be decided using DL concept subsumption.
  \item Based on the translation and the resulting description logic, we
    identify syntactic restrictions of SHACL for which the approach is sound and
    complete.
\end{enumerate}

\paragraph*{Organization} The paper first recalls the basic syntax and semantics
of SHACL and description logics in Section~\ref{sec:preliminaries}. We describe
how sets of SHACL shapes are translated into a DL knowledge base in
Section~\ref{sec:translation}. Section~\ref{sec:deciding_containment}
investigates how to use standard DL entailment for deciding shape containment.
Finally, we discuss related work in Section~\ref{sec:related_work} and summarize
our results.  \ifExtended{}{An extended version of this paper that includes full
proofs and additional explanations will be available on Arxiv.}

\end{document}

\section{Preliminaries}
\label{sec:preliminaries}

\subsection{Shape Constraint Language}
\label{sec:shacl}

The Shapes Constraint Language (SHACL) is a W3C standard for validating RDF
graphs. For this, SHACL distinguishes between the \emph{shape graph} that
contains the schematic definitions (e.\,g., Fig.~\ref{subfig:shape_graph_intro})
and the \emph{data graph} that is being validated (e.\,g.,
Fig.~\ref{subfig:data_graph_intro}). A shape graph consists of \emph{shapes}
that group \emph{constraints} and provide so called \emph{target nodes}. Target
nodes specify which nodes of the data graph have to be valid with respect to the
constraints in order for the graph to be valid. In the following, we rely on the
definitions presented by \cite{shaclSemantics}.

\paragraph*{Data graphs} We assume familiarity with RDF. We abstract away from
concrete RDF syntax though, representing an RDF Graph $G$ as a labeled oriented
graph $G = (V_G, E_G)$ where $V_G$ is the set of nodes of $G$ and $E_G$ is a set
of triples of the form $(v_1, p, v_2)$ meaning that there is an edge in $G$ from
$v_1$ to $v_2$ labeled with the property $p$. We use $\mathcal{V}$ to denote the
set of all possible graph nodes and $\mathcal{E}$ to denote the set of all
possible triples. A subset $\mathcal{V}_C \subseteq \mathcal{V}$ represents the
set of possible RDF classes. Furthermore, we use $\mathcal{G}$ to denote the set
of all possible RDF graphs.

\paragraph*{Constraints}

While shape graphs and constraints are typically given as RDF
graphs, we use a logical abstraction in the following. We use $\mathcal{N}_S$ to refer to the set
of all possible shape names. A constraint $\phi$ from the set of all
constraints $\Phi$ is then constructed as follows:
\begin{align}
  \phi ::= \; & \top \mid s \mid v \mid \phi_1 \land \phi_2 \mid \neg \phi \mid
                \geqslant_n\!\pe.\phi \\
  \pe ::= \; & p \mid \textasciicircum\pe \mid \pe_1 / \pe_2
\end{align}
where {$\top$} represents a constraint that is always true, $s \in \mathcal{N}_S$
references a shape name, $v \in \mathcal{V}$ is a graph node, $\neg \phi$
represents a negated constraint and $\geqslant_n \! \pe.\phi$ indicates that
there must be at least $n$ successors via the path expression $\pe$ that satisfy
the constraint $\phi$. For simplicity, we restrict ourselves to path expressions
$\pe$ comprising of either standard properties $p$, inverse of path $\textasciicircum\pe$,
and concatenations of two paths $\pe_1 / \pe_2$. We therefore leave out
operators for transitive closure and alternative paths. We use $\mathcal{P}$ to
indicate the set of all possible path expressions.
A number of additional syntactic constructs can be derived from these basic
constructors, including $\phi_1 \lor \phi_2$ for $\neg (\neg\phi_1 \land
\neg\phi_2)$, $\leqslant_n\!\pe.\phi$ for $\neg(\geqslant_{n+1}\!\pe.\phi)$,$=_n\!\pe.\phi$ for
$(\leqslant_n\!\pe.\phi) \land (\geqslant_n\!\pe.\phi)$, and $\forall\,\pe.\phi$ for
$\leq_0\!\pe.\neg\phi$. As an example, the constraint of
$\shapename{CubistShape}$ (see
Fig.~\ref{fig:example_validation_intro}) can be expressed as
$\geqslant_1\!(\textasciicircum\term{creator} / \term{style}).\term{cubism}$.

Evaluation of constraints is rather straightforward with the exception of
reference cycles. To highlight this issue, consider a shape name
$\shapename{Local}$ with its constraint
$\forall\,\term{knows}.\shapename{Local}$. In order to fulfill the constraint,
any graph node reachable through $\gedgenonodes{knows}$ must conform to
$\shapename{Local}$. Consider a graph with a single vertex \gnode{$b_1$} whose
\gedgenonodes{knows} property points to itself
\ifExtended{
(see Fig.~\ref{fig:shacl:rec_example}).

\begin{figure}[htp]
  \vspace{-22\in}
  \begin{subfigure}{0.5\textwidth}
    \[\phi_\term{Local} = \forall\,\term{knows}.\term{Local}\]
  \end{subfigure}
  \begin{subfigure}{0.5\textwidth}
    \centering
    \begin{tikzpicture}
      \node[iri] (b) {$b_1$}
        edge[arrin, loop right, min distance=10mm,in=0,out=90] node[label]{knows} (b);
    \end{tikzpicture}
  \end{subfigure}
  \caption{Illustration of a problematic, recursive case.}
  \label{fig:shacl:rec_example}
  \vspace{-42\in}
\end{figure}
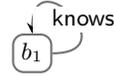
}{.}
\noindent Intuitively, there are two possible solutions. If \gnode{$b_1$} is
assumed to conform to $\shapename{Local}$, then the constraint is satisfied and
it is correct to say that \gnode{$b_1$} conforms to $\shapename{Local}$. If
\gnode{$b_1$} is assumed to not conform to $\shapename{Local}$, then the
constraint is violated and it is correct to say that \gnode{$b_1$} does not
conform to $\shapename{Local}$. We follow the proposal of~\cite{shaclSemantics}
and ground evaluation of constraints using \emph{assignments}. An assignment
$\sigma$ maps graph nodes $v$ to shape names $s$. Evaluation of constraints
takes an assignment as a parameter and evaluates the constraints with respect to
the given assignment. The case above is therefore represented through two
different assignments---one in which
$\shapename{Local}\in\sigma($\gnode{$b_1$}$)$ and a different one where
$\shapename{Local}\not\in\sigma($\gnode{$b_1$}$)$. We require assignments to be
total, meaning that they map all graph nodes to the set of all shapes that the
node supposedly conforms to. This disallows certain combinations of reference
cycles and negation in constraints, in essence requiring them to be stratified.
In contrast, \cite{shaclSemantics} also defines partial assignments, lifting
this restriction. Due to the lack of space, we refer to \cite{shaclSemantics}
for an in depth discussion on the differences of total and partial assignments.
\begin{definition}[Assignment]
  Let $G = (V_G,E_G)$ be an RDF graph and $S$ a set of shapes with its set of
  shape names $\names{S}$. An assignment $\sigma$ is a total function $\sigma :
  V_G \rightarrow 2^{\mathrm{Names}(S)}$ mapping graph nodes $v \in V_G$ to
  subsets of shape names. If a shape name $s \in \sigma(v)$, then $v$ is assigned
  to the shape name $s$. For all $s \not\in \sigma(v)$, the node $v$ is not
  assigned to the shape $s$.
\end{definition}
\noindent
Evaluating whether a node $v$ in $G$ satisfies a
constraint $\phi$, written $\evalc{\phi}$, is defined as shown in
Fig.~\ref{fig:evaluation:constraints}.
\begin{figure}[thp]
  \begin{subfigure}{0.3\textwidth}
    \begin{align*}
      \evalc{\top} &= \text{true}\\
      \evalc{\phi_1 \land \phi_2} &= \begin{cases}
        \text{true if } \evalc{\phi_1} = \text{true}~\land \\
        \hspace{0.25cm} \evalc{\phi_2} = \text{true}\\
        \text{false otherwise} \\
      \end{cases} \\
      \evalc{\geqslant_n\!\pe.\phi} &= \begin{cases}
        \text{true if } \vert\{v_2\mid (v_1,v_2) \in \evalp{r} \land \\
        \hspace{0.25cm} \evalc[v_2]{\phi} = \text{true} \}\vert \geq n \\
        \text{false otherwise} \\
      \end{cases}
%
    \end{align*}
  \end{subfigure}
  \hspace{-1cm}
  \begin{subfigure}{0.3\textwidth}
    \begin{align*}
      \evalc{\neg\phi} &= \begin{cases}
        \text{true if } \\
        \hspace{0.25cm} \evalc{\phi} = \text{false} \\
        \text{false otherwise} \\
      \end{cases} \\
      \evalc{v'} &= \begin{cases}
        \text{true if } v = v' \\
        \text{false otherwise} \\
      \end{cases} \\
      \evalc{s} &= \begin{cases}
        \text{true if } s \in \sigma(v) \\
        \text{false otherwise} \\
      \end{cases}
    \end{align*}
  \end{subfigure}
  \begin{subfigure}{1.0\textwidth}
    \begin{align*}
      \evalp{p} &= \{(v_1,v_2)\mid\!\exists\,p\!:\!(v_1,p,v_2)\in E_G\} \\
      \evalp{\textasciicircum \pe} &= \{ (v_2,v_1) \mid\!(v_1,v_2)\in\evalp{\pe} \} \\
      \evalp{\pe_1 / \pe_2} &= \{(v_1,v_2) \mid\!\exists\,v\!:\!(v_1,v) \in
      \evalp{\pe_1} \land (v,v_2) \in \evalp{\pe_2}\}
    \end{align*}
  \end{subfigure}
  \caption{Evaluation rules for constraints and path expressions.}
  \label{fig:evaluation:constraints}
\end{figure}

\paragraph*{Shapes and Validation}

A shape is modelled by a triple $(s, \phi, q)$. It consists of a shape name
$s$, a constraint $\phi$ and a query for target nodes $q$. Target nodes denote
those nodes which are expected to fulfill the constraint associated with the
shape. Queries for target nodes are built according to the following grammar:
\begin{align}
  q ::= \; & \bot \mid \{v_1, \ldots, v_n \} \mid \term{class}~v \mid \term{subjectsOf}~p \mid \term{objectsOf}~p
\end{align}
where $\bot$ represents a query that targets no nodes, $\{v_1 \ldots v_n \}$
targets all explicitly listed nodes with $v_1, \ldots, v_n \in \mathcal{V}$,
$\term{class}~v$ targets all instances of the class represented by $v$ where $v
\in \mathcal{V}_C$, $\term{subjectsOf}~p$ targets all subjects of the property $p$
and $\term{objectsOf}~p$ targets all objects of $p$. We use $\mathcal{Q}$ to
refer to the set of all possible queries and $\evalq{q}$ to denote the set of
nodes in the RDF graph $G$ targeted by the query $q$ (c.\,f.
Fig.~\ref{fig:evaluation:target_queries}).

\begin{figure}[tbp]
  \begin{subfigure}{0.4\textwidth}
    \begin{align*}
      \evalq{\bot} &= \emptyset \\
      \evalq{\{ v_1, \ldots, v_n \}} &= \{v_1, \ldots, v_n\}
    \end{align*}
  \end{subfigure}
  \begin{subfigure}{0.6\textwidth}
    \begin{align*}
      \evalq{\term{class}~v_2} &= \{v_1 \mid (v_1,\term{type},v_2) \in E_G\} \\
      \evalq{\term{subjectsOf}~p} &= \{v_1 \mid \exists v_2: (v_1,p,v_2) \in E_G \} \\
      \evalq{\term{objectsOf}~p}  &= \{v_2 \mid \exists v_1: (v_1,p,v_2) \in E_G\}
    \end{align*}
  \end{subfigure}
  \caption{Evaluation of target node queries.}
  \label{fig:evaluation:target_queries}
\end{figure}

A shape graph is then represented by a set of shapes $S$ whereas $\mathcal{S}$
represents the set of all possible sets of shapes. We assume for each
$(s,\phi,q) \in S$ that, if a shape name $s'$ appears in $\phi$, then there also
exists a $(s',\phi',q') \in S$. Similar to \cite{shaclSemantics}, we refer to
the language represented by the definitions above as $\mathcal{L}$. As an
example, Fig.~\ref{fig:shacl:set} shows the shape graph defined in
Fig.~\ref{subfig:shape_graph_intro} as a set of shapes.

\begin{figure}[ptb]
  \setlength{\abovedisplayskip}{0pt}
  \setlength{\belowdisplayskip}{0pt}
  \begin{alignat*}{2}
    S_1 = \{ \hphantom{aaaaaaaaa} \\
    (\shapename{PaintingShape},
    & \,\geqslant_1\!\term{exhibitedAt}.\top\,\land\,
    \forall\,\term{creator}.\term{PainterShape},
    & \,\term{class}~\term{Painting}), & \\
    (\shapename{PainterShape},\hphantom{i}
    & =_1\!\term{birthdate}.\top\,\land\,\forall\,\textasciicircum\term{creator}.\term{PaintingShape},
    & \,\bot), & \\
    (\shapename{CubistShape}\hphantom{iii},
    & \geqslant_1\!\textasciicircum\term{creator} / \term{style}.\term{cubism},
    & \,\bot)\hphantom{,} \\
    \} \hphantom{aaaaaaaaaaaaii}
  \end{alignat*}
%
  \caption{Representation of the shape graph shown in
    Fig.~\ref{subfig:shape_graph_intro} as a set of shapes.}
  \label{fig:shacl:set}
\end{figure}

Validating an RDF graph means finding a \emph{faithful assignment}. That is,
finding an assignment for which two conditions hold: First, if a node is a
target node of a shape, then the assignment must assign that shape to the node.
Second, if an assignment assigns a shape to a graph node, the constraint of the
shape must evaluate to true. Third, when a constraint evaluates to true (false)
on a node, that node must (not) be assigned to the corresponding shape. 
\begin{definition}[Faithful assignment]\label{def:faithful_assignment}
  An assignment $\sigma$ for a graph $G = (V_G, E_G)$ and a set of shapes $S$ is faithful,
  iff for each $(s, \phi, q) \in S$ and for each graph node $v \in V_G$, it
  holds that:
  \begin{itemize}
    \item $s \in \sigma(v) \Leftrightarrow \evalc{\phi}$.
    \item $v \in \evalq{q} \Rightarrow s \in \sigma(v)$.
  \end{itemize}
\end{definition}
%
\noindent A graph that is valid with respect to a set of shapes is said to
\emph{conform} to the set of shapes.

\begin{definition}[Conformance]\label{def:conformance}
  An RDF graph $G$ conforms to a set of shapes $S$ iff there is at least one
  faithful assignment $\sigma$ for $G$ and $S$. We write $\faith{G,S}$
  to denote the set of all faithful assignments for $G$ and $S$.
\end{definition}

\begin{figure}[htb]
  \setlength{\vgap}{0.6cm}
  \setlength{\hgap}{1.6cm}
  \centering
  \begin{tikzpicture}
    \node[iri] (Painting) {Painting};
    \node[iri, left=1.8cm of Painting] (Museum) {Museum};

    \node[iri, draw=highlight1, below=\vgap of Painting] (guernica) {guernica}
    edge[arrout] node[label]{type} (Painting);

    \node[iri, draw=highlight2, right=\hgap of guernica] (picasso) {picasso}
    edge[arrin] node[label]{creator} (guernica);


    \node[literal, right=\hgap of picasso] {``25.10.1881''}
    edge[arrin] node[label]{birthdate} (picasso);

    \node[iri, below=\vgap of Museum] (mncars) {mncars}
    edge[arrin] node[label]{exibitedAt} (guernica)
    edge[arrout] node[label]{type} (Museum);

    \node[iri, below=\vgap of guernica] (cubism) {cubism}
    edge[arrin] node[label]{style} (guernica);

    \node[] at (-4.5, -3) {\small $\sigma_1($};
    \node[smalliri] at (-4.0,-3) {};
    \node[] at (-1.35, -3) {\parbox{4.75cm}{\small $) = \emptyset$}};

    \node[] at (-2.5, -3) {\small $\sigma_1($};
    \node[smalliri,draw=highlight1] at (-2,-3) {};
    \node[] at (0.65, -3) {\parbox{4.75cm}{\small $) = \{\shapename{PaintingShape}\}$}};

    \node[] at (1.75, -3) {\small $\sigma_1($};
    \node[smalliri,draw=highlight2] at (2.25,-3) {};
    \node[] at (5, -3) {\parbox{5cm}{\small $) = \{\shapename{PainterShape}, \shapename{CubistShape}\}$}};
  \end{tikzpicture}
  \caption{Faithful assignment $\sigma_1$ for $S_1$ and the data graph shown in Fig.~\ref{subfig:data_graph_intro}.}
  \label{fig:faithful_assignment}
\end{figure}

\noindent
For the data graph shown in Fig.~\ref{subfig:data_graph_intro}, there is a
faithful assignment $\sigma_1$ that maps \shapename{PaintingShape} to
\gnode{guernica} and both \shapename{PainterShape} and \shapename{CubistShape}
to \gnode{picasso} (see Fig.~\ref{fig:faithful_assignment}). The assignment is
faithful because all instances of \gnode{Painting} are assigned to
\shapename{PaintingShape} and all nodes that are assigned to a shape satisfy the
constraints of the shape.

\end{document}

\subsection{Description Logics}
\label{subsec:description_logics}

\ifExtended{

In this paper, we focus on the highly-expressive DL $\mathcal{ALCOIQ}(o)$ as well as
decidable subsets of this logic.

\paragraph*{Concept Expressions}

\begin{figure}[tb]
  \centering
  \begin{tabular}{lccc} \toprule
    \multicolumn{2}{l}{\textbf{Constructor Name}} & \hphantom{aa}\textbf{Syntax}\hphantom{aa} & \textbf{Semantics}          \\
    \midrule
    atomic property              &  & $p$              & $p^I \subseteq \Delta^I \times \Delta^I$ \\
    inverse role                 &  & $r^{-}$          & $\{(o_2,o_1) \mid (o_1,o_2) \in r^I\}$         \\
    role composition             &  & $r_1 \circ r_2$  & $\{ (o_1,o_2) \mid (o_1,o) \in {r_1}^I \land (o,o_2) \in {r_2}^I \}$ \\
    \midrule
    atomic concept               &  & $A$            & $A^I \subseteq \Delta^I$  \\
    nominal concept              &  & $\{o_1, \ldots o_n\}$        & $\{ o_1^I, \ldots, o_n^I \}$               \\
    top                          &  & $\top$         & $\Delta^I$                \\
    negation                     &  & $\neg C$       &  $\Delta^I \setminus C^I$ \\
    conjunction                  &  & $C \sqcap D$   & $C^I \cap D^I$            \\
    qualified number restriction &  & $\geq_n\!r.C$ & $\{ o_1 \mid \vert \{ o_1 \mid (o_1,o_2) \in r^I \land o_2 \in C^I \} \vert \geq n  \}$ \\
    \bottomrule
  \end{tabular}
  \caption{Syntax and semantics of roles $r$ (above the line) and concept
    expressions $C, D$ (below the line).}
  \label{fig:dl:concept_expr}
\end{figure}

A description logic knowledge base $K$ is comprised of a set of axioms. Axioms
are either terminological statements, belonging to the so called TBox, or
assertional statements belonging to the ABox. Statements of a knowledge base $K$
are constructed using a signature $\mathrm{Sig}(K)$. The signature defines
atomic elements of the knowledge base, which can then be combined into
expressions using a range of available constructors. The signature of
a knowledge base $K$ is a triple $\mathit{Sig}(K) = (N_A,N_P,N_O)$ consisting of
a set of atomic concept names $N_A$ (e.\,g., \term{Painting}) that is a subset
of the set of all concept names $\mathcal{N}_A$, a set of relation or property
names $N_P$ (e.\,g.,~{\term{creator}}) which is a subset of $\mathcal{N}_P$ and
a set of atomic object names $N_O$ (e.\,g., \term{guernica}) which is a subset
of $\mathcal{N}_O$. Formal semantics of DL is based on first-order
logic~\cite{dlHandbook}. An interpretation $I$ is a pair consisting of
a non-empty universe $\Delta^I$ and an interpretation function $\cdot^I$. The
interpretation function $\cdot^I$ then maps each object $o \in N_O$ to an
element of the universe $o^I \in \Delta^I$. Furthermore, the interpretation $I$
assigns each atomic concept name $A \in N_A$ to a set $A^I \subseteq \Delta^I$
and each atomic property $p \in N_P$ to a binary relation $p^I \subseteq
\Delta^I \times \Delta^I$.

As shown in Fig.~\ref{fig:dl:concept_expr}, more complex expression can be built
from the atomic elements defined in the signature. Similar to path expressions,
role expressions, represented by the metavariable $r$, are either atomic
properties $p$, the inverse of a role expressions $r^-$, or the concatenation of
role expressions $r_1 \circ r_2$. Concept expressions, represented by the
metavariables $C$ and $D$, are either atomic concepts (represented by the
metavariable $A$), nominal concepts denoted by \term{$\{o\}$}, or \term{$\top$}
to represent the set of all objects. Concept expressions can also be composed
through negation \term{$\neg C$}, conjunction \term{$C \sqcap D$} or qualified
number restrictions \term{$\geq_n\!r.C$} expressing that there must be at least
$n$ successors via the role expression $r$ that belongs to the concept $C$.
Again, a number of additional constructors can be derived from the basic ones,
such as $\bot$ for $\neg\top$, and $\exists\,r.C$ for $\geq_1\!r.C$. We use
$\mathcal{C}$ to denote the set of all possible concept expressions and
$\mathcal{R}$ for the set of all possible role expressions.  As an example,
a concept expression representing the set of paintings that have
a creator~relation can be expressed as
$\term{Painting}\sqcap\exists\,\term{creator}.\top$.

\paragraph*{Axioms}

Using the previously defined concept expressions, terminological and assertional
axioms can be defined (see Figure~\ref{fig:dl:axioms}). Two concept expressions can be in a
subsumptive relationship {$C \sqsubseteq D$}. For example, the axiom
$\exists\,\term{exhibitedAt}.\top \sqsubseteq \term{Painting}$ says that everything that
has an \term{exhibitedAt}~property is also a \term{Painting}. Furthermore, we use {$C \equiv D$}
as a shorthand for the two axioms $C \sqsubseteq D$ and $D \sqsubseteq C$. In
terms of the assertional axioms, objects can be an instance of a concept
expression {$o:C$} (e.\,g. $\term{guernica}:\term{Painting}$), or be
connected to another object via a role expression {$(o_1,o_2):r$} (e.\,g.
$(\term{guernica},\term{picasso}):\term{creator}$).

\begin{figure}[tb]
\centering
\begin{tabular}{l p{2cm} p{2cm}} \toprule
    \textbf{Name} & \hfil \textbf{Syntax} & \hfil \textbf{Semantics}\\ \midrule
    concept inclusion   & \hfil $C \sqsubseteq D$ & \hfil $C^I \subseteq D^I$ \\ 
    concept assertion    & \hfil $o : C$           & \hfil $o^I \in C^I$       \\
    role assertion       & \hfil $(o_1,o_2):r$        & \hfil $(o_1^I,o_2^I) \in r^I$\\
    \bottomrule
\end{tabular}
\caption{Syntax and semantics of axioms.}
\label{fig:dl:axioms}
\end{figure}

\paragraph*{Entailment}

Logical entailment is defined through interpretations. In a given interpretation
$I$, a axiom is either true or false. For an axiom $\psi$ built
according to the syntax in Fig.~\ref{fig:dl:axioms}, we use the satisfaction
relationship $\models$ if its semantics constraint according to
Fig.~\ref{fig:dl:axioms} is true in a interpretation $I$ (written $I \models
\psi$). An interpretation $I$ satisfies a set of axioms $\Psi$ if
$\forall\,\psi \in \Psi: I \models \psi$. An interpretation
$I$ that satisfies all axioms of a knowledge base $K$, written $I \models K$, is
called a model of $K$. We use $\mo{K}$ to refer to the set of all models of $K$.
A axiom is entailed by a knowledge base $K$ if it is true in all
models of the knowledge base $K$.
\begin{definition}[Entailment]
  Let $K$ be a knowledge base, let $\psi$ refer to a axiom and let
  $\mathcal{I}$ be the set of all possible interpretations. The axiom
  $\psi$ is entailed by $K$, written $K \models \psi$, if $\forall\,I
  \in \mathcal{I}: (I \models K) \Rightarrow (I \models \psi)$.
\end{definition}

%
%
%
%

}
  {

We focus on the highly-expressive DL $\mathcal{ALCOIQ}(\circ)$ as well as
decidable subsets of this logic. We follow routine syntax and
interpretation-based semantics
(c.\,f.~\cite{introdl,dlWithComposition,dlHandbook}). $\mathrm{Sig}(K) =
(N_A,N_P,N_O)$ is the signature of a knowledge base $K$ comprising of a set of
atomic concept names $N_A$ that is a subset of the set of all possible atomic
concept names $\mathcal{N}_A$, a set of atomic property names $N_P$ (a
subset of $\mathcal{N}_P$) and a set of object names $N_O$ (a subset of
$\mathcal{N}_O$). From these, more complex role expressions, denoted by $r$, and 
concept expressions, denoted by $C$ and $D$, are built (see
Fig.~\ref{fig:dl:concept_expr}) whereby $\mathcal{C}$ denotes the set of all
possible concept expressions and $\mathcal{R}$ the set of all possible role
expressions.

\begin{figure}[tbp]
  \centering
  \begin{tabular}{lccc} \toprule
    \multicolumn{2}{l}{\textbf{Constructor Name}} & \hphantom{aa}\textbf{Syntax}\hphantom{aa} & \textbf{Semantics}          \\
    \midrule
    atomic property name         &  & $p$              & $p^I \subseteq \Delta^I \times \Delta^I$ \\
    inverse role                 &  & $r^{-}$          & $\{(o_2,o_1) \mid (o_1,o_2) \in r^I\}$         \\
    role composition             &  & $r_1 \circ r_2$  & $\{ (o_1,o_2) \mid (o_1,o) \in {r_1}^I \land (o,o_2) \in {r_2}^I \}$ \\
    \midrule
    atomic concept name          &  & $A$            & $A^I \subseteq \Delta^I$  \\
    nominal concept              &  & $\{o_1, \ldots o_n\}$        & $\{ o_1^I, \ldots, o_n^I \}$               \\
    top                          &  & $\top$         & $\Delta^I$                \\
    negation                     &  & $\neg C$       &  $\Delta^I \setminus C^I$ \\
    conjunction                  &  & $C \sqcap D$   & $C^I \cap D^I$            \\
    qualified number restriction &  & $\geq_n\!r.C$ & $\{ o_1 \mid \vert \{ o_1 \mid (o_1,o_2) \in r^I \land o_2 \in C^I \} \vert \geq n  \}$ \\
    \bottomrule
  \end{tabular}
  \caption{Syntax and semantics of roles $r$ (above the line) and concept
    expressions $C, D$ (below the line).}
  \label{fig:dl:concept_expr}
\end{figure}

\noindent
Axioms are either concept inclusions, concept assertions or role assertions (see
Fig.~\ref{fig:dl:axioms}). We furthermore use $C \equiv D$ as a shorthand for
the two axioms $C \sqsubseteq D$ and $D \sqsubseteq C$. In a given
interpretation $I = (\Delta, \cdot^I)$ comprised of a universe $\Delta$ and an
interpretation function $\cdot^I$, an axiom $\psi$ is either true or false. An
interpretation in which all axioms of $K$ are true is a model of $K$. We use
$\mathrm{Mod}(K)$ to denote the set of all models of $K$. An axiom $\psi$ is
entailed by $K$ written $K\models\psi$ if it is true in all models of $K$.
Lastly, we use $\mathcal{K}$ for the set of all possible knowledge bases.

\begin{figure}[tp]
\centering
\begin{tabular}{l p{2cm} p{2cm}} \toprule
    \textbf{Name} & \hfil \textbf{Syntax} & \hfil \textbf{Semantics}\\ \midrule
    concept inclusion   & \hfil $C \sqsubseteq D$ & \hfil $C^I \subseteq D^I$ \\
    concept assertion    & \hfil $o : C$           & \hfil $o^I \in C^I$       \\
    role assertion       & \hfil $(o_1,o_2):r$        & \hfil $(o_1^I,o_2^I) \in r^I$\\
    \bottomrule
\end{tabular}
\caption{Syntax and semantics of axioms.}
\label{fig:dl:axioms}
\end{figure}

%
%
%
%

}
\end{document}

\section{From SHACL Shape Containment to Description Logic Concept Subsumption}
\label{sec:translation}

Given two shapes $s$ and $s'$ that are elements of the same set of shapes $S$, we say
that $s$ is contained in $s'$ if any node that conforms to $s$ will also conform
to $s'$ for any given RDF data graph $G$ as well as any given faithful
assignment for $S$ and $G$.

\begin{definition}[Shape Containment]\label{def:containment}
  Let $S$ be a set of shapes with $s, s' \in \names{S}$.
  The shape $s$ is contained in shape $s'$ if:
  \[
    \forall\,G\in\mathcal{G}: \forall\,\sigma\in\faith{G,S}: \forall v\in V_G:
    s\in\sigma(v)\Rightarrow s'\in\sigma(v)\,\text{with}~s,s'\in\names{S}
  \]
  We use $\contained{s}{s'}$ to indicate that shape $s$ is contained in $s'$
  with respect to $S$.
\end{definition}

Both SHACL and description logics use syntactic formulas inspired by
first-order logic. However, their semantics are fundamentally different. For
SHACL, we follow the Datalog-inspired semantics introduced
by~\cite{shaclSemantics}. Description logics on the other hand adopt Tarskian-style 
semantics.   
To decide shape containment, we map sets of shapes syntactically into
description logic knowledge bases such that the difference in semantics can be
overcome. 

\begin{figure}[htb]
  \centering
  \begin{tikzpicture}
    \node[component] (shapes) {Shape};
    \node[component,right=4.5cm of shapes] (axioms) {Axioms}
    edge[rolein, bend right=10] node[label]{\tiny $\shapesToKB$} (shapes);

    \node[component, below=1cm of shapes] (constraints) {Constraint}
    edge[built] (shapes);
    \node[component, below right=0.4cm and 1cm of shapes, align=center]
      (targetnodes) {Target Node \\ Query};

    \draw[rounded corners=5pt, built] (targetnodes) -- +(0,1) -- (shapes);

    \node[component, below left=3.75cm and 0.3cm of shapes] (shapenames) {Shape
      \\ Name};

    \draw[rounded corners=5pt, built] (shapenames) -- +(0,4) -- (shapes);


    \draw[rounded corners=5pt, built] (shapenames) -- (-1.3,-1.45) -- (constraints);

    \node[component, below=0.4cm of axioms] (conceptexpr) {Concept \\ Expression}
    edge[rolein] node[label, xshift=-0.1cm]{\tiny $\targetNoToConceptExpr$} (targetnodes)
    edge[rolein, bend left=10] node[label,xshift=0.5cm]{\tiny $\constrToConceptExpr$} (constraints)
    edge[built] (axioms);

    \node[component, below=0.4cm of constraints] (pathExpr) {Path \\ Expression}
    edge[built] (constraints);

    \node[component, below right=0.75cm and 0.1cm of constraints] (graphNode) {Graph \\ Node};

    \draw[rounded corners=5pt, built] (graphNode) -- +(1,0) -- (targetnodes);
    \draw[rounded corners=5pt, built] (graphNode) -- (1.25,-1.75) -- (constraints);

    \node[component, below=0.65cm of pathExpr] (property) {RDF property}
    edge[built] (pathExpr);

    \node[component, below=2.8cm of targetnodes,xshift=-0.2] (classes) {RDF class}
    edge[built] (targetnodes);

    \node[component, below=0.75cm of conceptexpr] (roleexpr) {Role \\ Expression}
    edge[built] (conceptexpr)
    edge[rolein, bend right=10] node[label,xshift=0.5cm]{\tiny $\pathToRole$} (pathExpr);

    \node[component, below left=-0.35cm and 0.25cm of roleexpr,yshift=0.1cm] (objectNames) {Object \\ Name}
    edge[rolein,bend left=10] node[label,xshift=0.1cm]{\tiny $\toObjectName$} (graphNode);

    \draw[rounded corners=5pt, built] (objectNames) -- (4.1,-1.5) -- (conceptexpr);

    \node[component, below=0.5cm of roleexpr] (propertyName) {Atomic Property \\ Name}
    edge[built] (roleexpr)
    edge[rolein] node[label,xshift=0.1cm] {\tiny $\toAtomicConcept$} (property);

   \node[component, below right=1.15cm and 0.5cm of roleexpr] (conceptName) {Atomic Concept \\ Name}
   edge[rolein,bend left=10] node[label]{\tiny $\toAtomicConcept$} (shapenames)
   edge[rolein] node[label]{\tiny $\toAtomicConcept$} (classes);

   \draw[rounded corners=5pt, built] (conceptName) -- (7.65,-2) -- (conceptexpr);
%
%
%
%
   \matrix [draw=black,
   right=0.75cm of axioms] {
      \node[] (m) at (0,0){}; \draw (m) edge[built,-latex] +(0.75,0); & \node[]{\tiny built using}; \\
      \node[] (m) at (0,0){}; \draw (m) edge[roleout] +(0.75,0); & \node[]{\tiny maps to}; \\
      \\
    };
  \end{tikzpicture}
  \caption{Syntactic translation of SHACL to description logics.}
  \label{fig:translation_functions}
\end{figure}
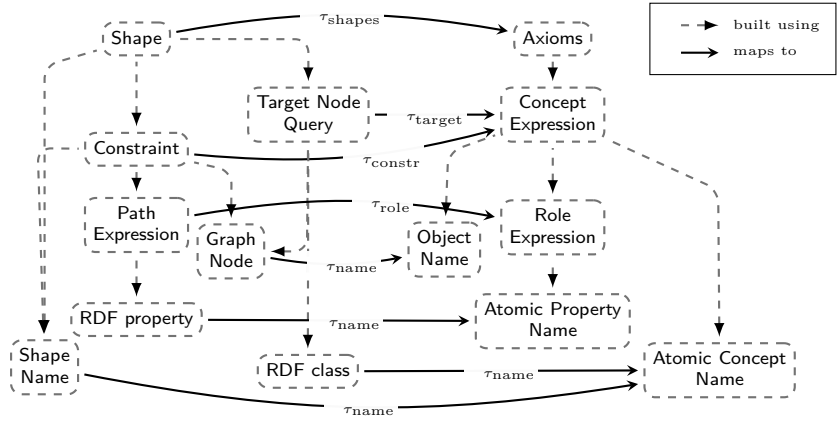

The function $\shapesToKB$ maps a set of
shapes $S$ to a description logic knowledge base $\constructed{K}{S}$ using four
auxiliary functions (see Fig.~\ref{fig:translation_functions}): First,
$\toAtomicConcept$ maps shape names, RDF classes as well as properties and graph
nodes onto atomic concept names, atomic property names and object names. Second,
$\pathToRole$ maps SHACL path expressions to DL role expressions.  Third,
$\constrToConceptExpr$ maps constraints to concept expressions. Fourth,
$\targetNoToConceptExpr$ maps queries for target nodes to concept expressions.
The function $\shapesToKB$ maps a set of shapes $S$ to a set of axioms such that
$\contained{s}{s'}$ is true if $\constructed{K}{S} \models \toAtomicConcept[s]
\sqsubseteq \toAtomicConcept[s']$.

To prove this property, we show that every finite model of $\constructed{K}{S}$
can be used to construct an RDF graph $G$ and an assignment that is faithful
with respect to $G$ and $S$. Likewise, a model of $\constructed{K}{S}$ can be
constructed from an assignment that is faithful with respect to $S$ and any
given RDF graph $G$.

\subsection{Syntactic Mapping}

We map the set of shapes $S$ into a knowledge base $\constructed{K}{S}$ by
constraints and target node queries of each shape using the functions
$\pathToRole$, $\constrToConceptExpr$, $\targetNoToConceptExpr$, and
$\shapesToKB$. All those functions rely on $\toAtomicConcept$ which maps atomic
elements used in SHACL to atomic elements of a DL knowledge base:
\begin{definition}[Mapping atomic elements]
  The function
  $\toAtomicConcept:\mathcal{N}_S\cup\mathcal{V}_C\cup\mathcal{V}\cup\mathcal{E}\rightarrow\mathcal{N}_A\cup\mathcal{N}_P\cup\mathcal{N}_O$
  is an injective function mapping shape names and RDF classes onto atomic concept
  names, graph nodes onto object names as well as properties onto atomic property
  names.
\end{definition}
\begin{definition}[Mapping path expressions to DL roles]
  The \emph{path mapping} function $\pathToRole: \mathcal{P} \rightarrow
  \mathcal{R}$, is defined as follows:

  \begin{tabular}{lcl}
    $\pathToRole(p)$ & $=$ & $\toObjectName(p)$ \\
    $\pathToRole(\textasciicircum\rho)$ & $=$ & $\pathToRole(\rho)^-$ \hfill \\
    $\pathToRole(\rho_1 / \rho_2)$ & $=$ & $\pathToRole(\rho_1) \circ
    \pathToRole(\rho_2)$
  \end{tabular}
\end{definition}
\begin{definition}[Mapping constraints to DL concept expressions]
  \label{def:containment:constraint2concept}
  The \emph{constraint mapping} $\constrToConceptExpr: \Phi \rightarrow
  \mathcal{C}$ is defined as follows:

  \begin{tabular}{lcl}
    $\constrToConceptExpr(\top)$ & $=$ & $\top$ \\
    $\constrToConceptExpr(s)$ & $=$ & $\toAtomicConcept[s]$ \hfill \\
    $\constrToConceptExpr(v)$ & $=$ & $\{ \toObjectName[v] \}$ \\
    $\constrToConceptExpr(\phi_1 \land \phi_2)$ & $=$ &
    $\constrToConceptExpr(\phi_1) \sqcap \constrToConceptExpr(\phi_2)$ \\
    $\constrToConceptExpr(\neg \phi)$ & $=$ & $\neg\constrToConceptExpr(\phi)$ \\
    $\constrToConceptExpr(\geqslant_n\!\pe.\phi)$ & $=$ &
    $\geq_n\!\pathToRole(\pe).\constrToConceptExpr(\phi)$
  \end{tabular}
\end{definition}

\begin{definition}[Mapping target node queries to DL concept
  expressions]\label{def:targets2concepts}
  The \emph{target node mapping} $\targetNoToConceptExpr : \mathcal{Q} \rightarrow
  \mathcal{C}$ is defined as follows:

  \begin{tabular}{lclr}
    $\targetNoToConceptExpr(\bot)$ & $=$ & $\bot$ & \\
    $\targetNoToConceptExpr(\{v_1, \ldots, v_n\})$ & $=$ & $\{\toObjectName[v_1], \ldots, \toObjectName[v_n]\}$ & \\
    $\targetNoToConceptExpr(\mathtt{class}~v)$ & $=$ & $\toAtomicConcept[v]$ &  \\
    $\targetNoToConceptExpr(\mathtt{subjectsOf}~p)$ & $=$
                                                    & $\exists\,\toAtomicConcept(p).\top$ \\
    $\targetNoToConceptExpr(\mathtt{objectsOf}~p)$ & $=$
                                                   & $\exists\,\toAtomicConcept(p)^{-}.\top$ \\
  \end{tabular}
\end{definition}
\noindent
The mapping $\targetNoToConceptExpr(q)$ of a target query $q$ is defined such
that querying for the instances of $q$ returns exactly the same nodes from the
data graph. Likewise, the mapping $\constrToConceptExpr(\phi)$ is defined such
that it contains those nodes for which $\phi$ evaluates to true and
$\pathToRole$ that the interpretation of the role expression contains those
nodes that are also in the evaluation of the path expression.
%
%
$\shapesToKB$ generalizes the construction to sets of shapes:
\begin{definition}[Mapping sets of shapes to DL axioms]\label{th:shapesToKB}
  The \emph{shape mapping} function $\shapesToKB : \mathcal{S} \rightarrow
  \mathcal{K}$ is defined as follows:
  \begin{align*}
    \shapesToKB(S) = \bigcup_{(s,\phi,q)\in S} \{
     \targetNoToConceptExpr(q) \sqsubseteq \toAtomicConcept[s], 
     \constrToConceptExpr(\phi) \equiv \toAtomicConcept[s]
    \}
  \end{align*}
\end{definition}
\noindent
%
To illustrate the function
$\shapesToKB$, the translation of the set of shapes $\shapesToKB(S_1) = \constructed{K}{S_1}$
is shown in Fig.~\ref{fig:translation:s1}.
\begin{figure}[htp]
  \setlength{\abovedisplayskip}{0pt}
  \setlength{\belowdisplayskip}{0pt}
  \setlength{\abovedisplayshortskip}{0pt}
  \setlength{\belowdisplayshortskip}{0pt}
  \begin{alignat*}{3}
    \constructed{K}{S_1} = \{\, & \term{Painting} \sqsubseteq \term{PaintingShape}, & \\
    & \geq_1\!\term{exhibitedAt}.\top \, \sqcap
    \forall\,\term{creator}.\term{PainterShape} \equiv \term{PaintingShape}, & \\
    & \,\bot \sqsubseteq \term{PainterShape}, & \\
    & \geq_1\!\term{birthdate}.\top \,
    \sqcap\,\forall\,\term{creator}^{-}.\term{PaintingShape} \equiv
    \term{PainterShape}, & \\
    & \bot \sqsubseteq \term{Cubist}, & \\
    & \geq_1\!\term{creator}^{-}\circ\term{style}.\term{\{cubism\}} \equiv \term{CubistShape} & \}
  \end{alignat*}
  \caption{Translation $\shapesToKB(S_1) = \constructed{K}{S_1}$ of the set of shapes $S_1$.}
  \label{fig:translation:s1}
\end{figure}

%
%

\subsection{Construction of Faithful Assignments and Models}

%
Given our translation, we now show that the notion of faithful assignments of
SHACL and \emph{finite models} in description logics coincide.

\begin{definition}[Finite model]
  Let $K$ be a knowledge base and $I \in \mo{K}$ a model of $K$. The model $I$
  is finite, if its universe $\Delta^I$ is finite~\cite{finiteModel}. We use
  $\finmo{K}$ to refer to the set of all finite models of $K$.
\end{definition}
\noindent
Given an RDF data graph $G$, a set of shapes $S$ and an assignment $\sigma$ that
is faithful with respect to $S$ and $G$, we construct an interpretation
$\constructed{I}{G,\sigma}$ that is a finite model for the knowledge base
$\constructed{K}{S}$.
%
\begin{definition}[Construction of the finite model $\constructed{I}{G,\sigma}$]
  \label{def:containment:constructed_interpretation}
  Let $S$ be a set of shapes, $G = (V_G, E_G)$ an RDF data graph and $\sigma$ an
  assignment that is faithful with respect to $S$ and $G$. Furthermore, let
  $\toGraphNode$ be the inverse of the function $\toAtomicConcept$. The finite
  model $\constructed{I}{G,\sigma}$ for the knowledge base $\shapesToKB(S) =
  \constructed{K}{S}$ is constructed as follows:
  \begin{enumerate}
    \item All objects are interpreted as themselves: $\forall\,o\in N_O: o^I = o$.
    \item A pair of objects is contained in the interpretation of a relation if
      the two objects are connected in the RDF data graph:\\
      $\forall\,p \in N_P: \forall
        \,o_1,o_2 \in N_O:
        (o_1^{\constructed{I}{G,\sigma}},o_2^{\constructed{I}{G,\sigma}}) \in
        p^{\constructed{I}{G,\sigma}}
        \text{if}~(\toGraphNode[o_1],p,\toGraphNode[o_2]) \in
        (E_G\setminus\{(v_1,\mathtt{type},v_2) \in E_G\})$.
  \item Objects are in the interpretation of a concept if this concept is
    a class used in the RDF data graph and the object is an instance of this class
    according to the graph:\\
    $\forall A_v \in N_A: \forall o \in N_O:  o^{\constructed{I}{G,\sigma}} \in A_v^{\constructed{I}{G,\sigma}}~\text{if}~(\toGraphNode[o],\mathtt{type},\toGraphNode[A_v])\in E_G$.
  \item Objects are in the interpretation of a concept if the concept is a shape
    name and the assignment $\sigma$ assigns the shape to the object:\\
    $\forall A_s \in N_A: \forall o \in N_O: o^{\constructed{I}{G,\sigma}} \in A_s^{\constructed{I}{G,\sigma}}~\text{if}~\toGraphNode(A_s)\in\sigma(\toGraphNode[o])$.
  \end{enumerate}
\end{definition}
\noindent
The interpretation $\constructed{I}{G,\sigma}$ is a model of the knowledge base
$\constructed{K}{S}$. Before we show this, it is important to notice that the
interpretation of role expressions constructed through $\pathToRole$ contains
the same nodes in the interpretation $\constructed{I}{G,\sigma}$ as the
evaluation of the path expression.
\begin{lemma}
  Let $S$ be a set of shapes, $G$ an RDF data graph and $\sigma$ an assignment
  that is faithful with respect to $S$ and $G$. Furthermore, let
  $\constructed{I}{G, \sigma}$ be an interpretation for $\constructed{K}{S}$.
  It holds that $\forall (o_1, o_2)\in \pathToRole(\pe)^{\constructed{I}{G,\sigma}} \Rightarrow
  (\toGraphNode(o_1),\toGraphNode(o_2)) \in \evalp{\pe}$ for any path expression
  $\pe$.
\end{lemma}

\begin{proof}
  The interpretation $\constructed{I}{G,\sigma}$ contains all properties of the
  RDF graph. The result is then immediate from the evaluation rules of path expressions
  (c.\,f.~Fig.~\ref{fig:evaluation:constraints}) and semantics of role
  expressions (c.\,f.~Fig.~\ref{fig:dl:concept_expr}).
\end{proof}

\begin{theorem}\label{th:constructed_interpretation}
  Let $S$ be a set of shapes, $G$ an RDF data graph and $\sigma$ an assignment
  that is faithful with respect to $S$ and $G$. Furthermore, let
  $\constructed{K}{S}$ be a knowledge base that is constructed through
  $\shapesToKB(S)$. The interpretation $\constructed{I}{G,\sigma}$ is a finite
  model of $\constructed{K}{S}$ ($\constructed{I}{G,\sigma} \models
  \constructed{K}{S}$).
\end{theorem}

\ifExtended{
\begin{proof}
  $\constructed{I}{G,\sigma}$ is a finite model of $\constructed{K}{S}$ iff
  $\forall \psi \in \constructed{K}{S}: \constructed{I}{G,\sigma} \models \psi$
  and $\Delta^{\constructed{I}{G,\sigma}}$ is finite. For each shape $(s,\phi,q)
  \in S$, there are two axioms in $\constructed{K}{S}$. First, the axiom
  $\targetNoToConceptExpr(q) \sqsubseteq \toAtomicConcept[s]$.  Second, the
  axiom $\constrToConceptExpr(\phi)\equiv\toAtomicConcept[s]$.
  $\constructed{I}{G,\sigma}$ must satisfy both axioms.
  We start by noting that $\Delta^{\constructed{I}{G,\sigma}}$ is finite since
  the RDF graph $G$ from which the set of objects $N_O$ is constructed is
  finite.
  The axiom {$\targetNoToConceptExpr(q) \sqsubseteq
  \toAtomicConcept[s]$} is satisfied in $\constructed{I}{G,\sigma}$ by examining each case of
  $\targetNoToConceptExpr$ individually:
  \begin{description}
    \item[$\targetNoToConceptExpr(\bot) = \bot$] Vacuously satisfied as
      {$\bot$} is a subset of every concept expression.

    \item[$\targetNoToConceptExpr(\{\overline{v}\}) = \{\toObjectName(\overline{v})\}$] Target
      nodes consist of an enumeration of nodes. $\sigma$ is only faithful if the
      shape $s$ is assigned to all those nodes. Likewise,
      {$\{\toObjectName[\overline{v}]\}$}
      constitutes a concept expression that is an enumeration of graph nodes.
      As all nodes that are assigned to $s$ in $\sigma$ are also in the
      interpretation {$\toAtomicConcept[s]^{\constructed{I}{G,\sigma}}$} of
      {$\toAtomicConcept[s]$}, the axiom {$\{\toObjectName(\overline{v})\}
      \sqsubseteq \toAtomicConcept[s]$} must be satisfied in $\constructed{I}{G,\sigma}$.

    \item[$\targetNoToConceptExpr(\term{class}~v)
      = \toAtomicConcept(v)$] The assignment $\sigma$ is only faithful if the
      shape $s$ is assigned to all instances of {$v$}. Due to the
      construction of $\constructed{I}{G,\sigma}$, all instances
      of {$v$} are in the interpretation
      {${\toAtomicConcept[s]}^{\constructed{I}{G,\sigma}}$} of {$\toAtomicConcept[s]$}.
      Subsequently, {$\toAtomicConcept(v) \sqsubseteq \toAtomicConcept[s]$} must
      be satisfied in $\constructed{I}{G,\sigma}$.

    \item[$\targetNoToConceptExpr(\term{subjectsOf}~p)
      = \exists\,\pathToRole(p).\top$] The assignment $\sigma$ is faithful if
      shape $s$ is assigned to all nodes that have the given property
      $p$. Since the interpretation $\constructed{I}{G,\sigma}$ is
      constructed using $\sigma$, all nodes having that property must be in the
      interpretation ${\toAtomicConcept[s]}^{\constructed{I}{G,\sigma}}$ of
      {$\toAtomicConcept[s]$}.
      Subsequently, $\exists\,\pathToRole(p).\top \sqsubseteq
      \toAtomicConcept[s]$ must be satisfied in $\constructed{I}{G,\sigma}$.

    \item[$\targetNoToConceptExpr(\term{objectsOf}~p)
      = \exists\,p^-.\top$] The assignment $\sigma$ is faithful
      if shape $s$ is assigned to all nodes that have the given incoming
      $p$ relation. Due to the construction of $\constructed{I}{G,\sigma}$
      (c.\,f.~Definition~\ref{def:containment:constructed_interpretation}), all
      nodes that have an incoming relation via the property must be in the
      interpretation ${\toAtomicConcept[s]}^{\constructed{I}{G,\sigma}}$ of {$\toAtomicConcept[s]$}. Subsequently, the axiom
      {$\exists\,p^-.\top \sqsubseteq \toAtomicConcept[s]$} must
      be satisfied in $\constructed{I}{G,\sigma}$.
  \end{description}

  \noindent
  We continue by showing that {$\constrToConceptExpr(\phi) \equiv
  \toAtomicConcept[s]$} is satisfied in $\constructed{I}{G,\sigma}$ via induction over
  $\constrToConceptExpr$:
  \begin{description}
    \item[$\constrToConceptExpr(\top) = \top$] If {$\phi = \top$}, then
      {$\llbracket \top \rrbracket^{o,G,\sigma}$} evaluates to true for all nodes.
      Therefore the shape $s$ is assigned to all nodes and subsequently the
      concept {${\toAtomicConcept[s]}^{\constructed{I}{G,\sigma}}$} contains all nodes due to the construction of
      $\constructed{I}{G,\sigma}$. This is equivalent to the concept {$\top$}.

    \item[$\constrToConceptExpr(s') = {\toAtomicConcept[s']}$] If
      {$\phi = s'$},
      then {$\llbracket s' \rrbracket^{o,G,\sigma}$} evaluates to true if $s' \in
      \sigma(o)$. Due to the construction of $\constructed{I}{G,\sigma}$, all nodes for which
      this is true must also be in the interpretation
      {${\toAtomicConcept[s]}^{\constructed{I}{G,\sigma}}$} of
      {$\toAtomicConcept[s]$}. Subsequently, the axiom
      {$\toAtomicConcept[s'] \equiv \toAtomicConcept[s]$} must be true in
      $\constructed{I}{G,\sigma}$.

    \item[$\constrToConceptExpr(v) = \{\toObjectName(v)\}$] If {$\phi = v$}, then {$\llbracket v
      \rrbracket^{v',G,\sigma}$} evaluates to true if $v = v'$. Therefore, the
      shape $s$ is assigned to this node. Due to the construction of
      $\constructed{I}{G,\sigma}$, $\toObjectName(v)$ is the only node in the interpretation of
      {${\toAtomicConcept[s]}^{\constructed{I}{G,\sigma}}$}. The axiom
      {$\{\toObjectName(v)\}\equiv \toAtomicConcept[s]$} must therefore be true in
      $\constructed{I}{G,\sigma}$.

    \item[$\constrToConceptExpr(\phi_1 \land \phi_2)
      = \constrToConceptExpr(\phi_1) \sqcap \constrToConceptExpr(\phi_2)$]
      Evaluation of the constraint {$\llbracket \phi_1 \land \phi_2
      \rrbracket^{o,G,\sigma}$} evaluates to true for nodes where
      $\phi_1$ and $\phi_2$ evaluate to true. By induction hypothesis,
      {$\constrToConceptExpr(\phi_1) = C$} is a concept expression that is
      equivalent to the set of nodes for which $\phi_1$ evaluates to true.
      Likewise for {$\constrToConceptExpr(\phi_2) = D$}. The set of nodes for
      which both $\phi_1$ and $\phi_2$ evaluate to true must therefore be the
      intersection of {$C \sqcap D$}. Due to the construction of $\constructed{I}{G,\sigma}$,
      those nodes must also be in the interpretation
      {${\toAtomicConcept[s]}^{\constructed{I}{G,\sigma}}$} of
      {$\toAtomicConcept[s]$}.
      The axiom must therefore be true.

    \item[$\constrToConceptExpr(\neg \phi_1) = \neg
      \constrToConceptExpr(\phi_1)$] By hypothesis,
      {$\constrToConceptExpr(\phi_1) = C$} is a concept expression that is
      equivalent to the set of nodes for which $\phi_1$ evaluates to true.
      Evaluation of the constraint {$\llbracket \neg \phi_1
      \rrbracket^{o,G,\sigma}$} evaluates to true for those nodes in which
      $\phi_1$ evaluates to false. Since those nodes are assigned to $s$ in
      $\sigma$, the interpretation
      {${\toAtomicConcept[s]}^{\constructed{I}{G,\sigma}}$} of
      {$\toAtomicConcept[s]$} must also be those
      nodes. This is equivalent to the interpretation of {$\neg C$}. The axiom
      {$\neg C \equiv \toAtomicConcept[s]$} must therefore be satisfied in $\constructed{I}{G,\sigma}$.



    \item[$\constrToConceptExpr(\geqslant_n\!\pe.\phi_1) = \geq_n\!r.\constrToConceptExpr(\phi_1)$]
      By hypothesis, {$\constrToConceptExpr(\phi_1) = C$} is a concept
      expression that represents the set of graph nodes for which $\phi_1$
      evaluates to true. {$\llbracket \geqslant_n\!\pe.\phi_1
      \rrbracket^{o,G,\sigma}$}
      evaluates to true for those nodes that have at least $n$ successors via
      $\pe$ and for which $\phi_1$ evaluates to true. Those nodes must also be in
      the interpretation {${\toAtomicConcept[s]}^{\constructed{I}{G,\sigma}}$} of
      {$\toAtomicConcept[s]$}. Due to the construction of
      $\constructed{I}{G,\sigma}$, all graph nodes having $n$ successor via $r$ in $G$ must
      also have $n$ successors in $\constructed{I}{G,\sigma}$. Subsequently, the axiom
      {$\geq_n\!r.C \equiv \toAtomicConcept(s)$} must be true in $\constructed{I}{G,\sigma}$. \qedhere
  \end{description}
\end{proof}
}{
  \begin{proof}[Proof (Sketch)]
    $\constructed{I}{G,\sigma}$ is finite because the RDF graph $G$ has only a
    finite number of graph nodes. Furthermore, $\constructed{I}{G,\sigma}$ satisfying the
    axioms created by $\shapesToKB$ can be shown via induction over the
    mapping rules for $\constrToConceptExpr$ and $\targetNoToConceptExpr$.
  \end{proof}
}
\noindent
Furthermore, we show that any finite model $I$ of a knowledge base $\constructed{K}{S}$
built from a set of shapes $S$ can be transformed into an RDF graph $\constructed{G}{I}$
and an assignment $\constructed{\sigma}{I}$ such that $\constructed{\sigma}{I}$ is faithful with respect to
$S$ and $\constructed{G}{I}$. We construct $\constructed{G}{I}$ and $\constructed{\sigma}{I}$ in the following manner:
\begin{definition}[Construction of $\constructed{G}{I}$ and $\constructed{\sigma}{I}$]
\label{def:containment:construction_assignment}
  Let $S$ be a set of shapes and $\constructed{K}{S}$ a knowledge base
  constructed via $\shapesToKB(S)$. Furthermore, let $I \in \finmo{\constructed{K}{S}}$ be a finite model of
  $\constructed{K}{S}$. The RDF graph $\constructed{G}{I} =
  (V_\constructed{G}{I}, E_\constructed{G}{I})$ and the assignment
  $\constructed{\sigma}{I}$ can then be constructed as follows:
  \begin{enumerate}
    \item The interpretations of all relations are interpreted as relations
      between graph nodes in the RDF graph:\\
      \hphantom{a}$\forall p \in N_P: (o^I,o'^I)\in
      p^I \Rightarrow (\toGraphNode(o),p,\toGraphNode(o')) \in E_\constructed{G}{I}$.
    \item The interpretations of all concepts that are not shape names are
      triples indicating an instance in the RDF graph: \\
      \hphantom{a}$\forall A \in N_A: (o^I
      \in A^I \land A \not\in \names{S}) \Rightarrow (\toGraphNode(o),\mathtt{type},\toGraphNode(A)) \in
      E_\constructed{G}{I}$.
    \item The interpretations of all concept names that are shape names are used
      to construct the assignment \\
      \hphantom{a}$\constructed{\sigma}{I}$: $\forall A \in N_A: (o^I \in A^I
      \land A \in \names{S}) \Rightarrow \toGraphNode(A) \in \constructed{\sigma}{I}(\toGraphNode(o))$.
  \end{enumerate}
\end{definition}

An assignment $\constructed{\sigma}{I}$ constructed in this manner is faithful with respect to
the constructed RDF graph $\constructed{G}{I}$ and the set of shapes $S$.

\begin{theorem}\label{th:constructed_assignment}
  Let $S$ be a set of shapes and $\constructed{K}{S}$ be a knowledge base
  constructed through $\shapesToKB(S)$. Furthermore, let $I \in
  \finmo{\constructed{K}{S}}$ be a finite model for $\constructed{K}{S}$. The
  assignment $\constructed{\sigma}{I}$ is faithful with respect to $S$ and
  $\constructed{G}{I}$.
\end{theorem}

\ifExtended{
\begin{proof}
  $\constructed{\sigma}{I}$ is faithful with respect to $S$ and $\constructed{G}{I}$ if two conditions hold:
  \begin{enumerate}
    \item Each shape is assigned to all of its target nodes.
    \item If a shape is assigned to a node, then the constraint evaluates to
      true. Likewise, if the constraint evaluates to true for a node, then the shape is
      assigned to the node.
  \end{enumerate}
  The knowledge base $\shapesToKB(S) = \constructed{K}{S}$ contains two axioms of the form
  {$\targetNoToConceptExpr(q) \sqsubseteq \toAtomicConcept[s]$} and
  {$\constrToConceptExpr(\phi) \equiv \toAtomicConcept[s]$} for each $(s,\phi,q)$.
  $I \in \mo{\constructed{K}{S}}$ satisfies those axioms. We proceed by examining each case of
  $\targetNoToConceptExpr$ individually:
  \begin{description}
    \item[$\targetNoToConceptExpr(\bot) = \bot$] Vacuously satisfied as no
      target nodes exist.

    \item[$\targetNoToConceptExpr(\{v_1, \ldots, v_n\})
      = \{\toObjectName(v_1), \ldots, \toObjectName(v_n)\}$]
      The target nodes are an enumeration of nodes $v_1, \ldots, v_n$. $I$ is
      only a model if $\{\toObjectName(v_1), \ldots, \toObjectName(v_n)\}^I \subseteq {\toAtomicConcept[s]}^I$ is true. Due to the
      construction of $\constructed{G}{I}$, all nodes $\toObjectName(v_1), \ldots, \toObjectName(v_n)$ must exist in $\constructed{G}{I}$. Due
      to the construction of $\constructed{\sigma}{I}$, the shape $s$ is assigned to all nodes
      in {${\toAtomicConcept[s]}^I$}. As such, the shape is assigned to all of its target nodes.

    \item[$\targetNoToConceptExpr(\term{class}~v)
      = \toAtomicConcept(v)$] Target nodes are instances of a concept. $I$ is
      only a model if $\toAtomicConcept(v)^I \subseteq
      {\toAtomicConcept[s]}^I$ is true. Due to
      the construction of $\constructed{G}{I}$, the concept {$\toAtomicConcept(v)$} and
      its instances $\toAtomicConcept(v)^I$ must exist in $\constructed{G}{I}$. Due to the
        construction of $\constructed{\sigma}{I}$, $s$ is assigned to all nodes in
        $\toAtomicConcept(v)^I$. As such, the shape is assigned to all of its
        target nodes.

    \item[$\targetNoToConceptExpr(\term{subjectsOf}~p)
      = \exists\,p.\top$] Target nodes are all subjects of a property. $I$ is
      only a model if {${\exists\,p.\top}^I \subseteq {\toAtomicConcept[s]}^I$}
      is true. Due to the construction of $\constructed{G}{I}$, all nodes in
      {${\exists\,p.\top}^I$} must exist in $\constructed{G}{I}$ and have the property. Due to
      the construction of $\constructed{\sigma}{I}$, the shape $s$ is assigned to all nodes in
      {${\exists\,p.\top}^I$}. As such, the shape is assigned to all of its
      target nodes.

    \item[$\targetNoToConceptExpr(\term{objectsOf}~p)
      = \exists\,p^-.\top$] Target nodes are objects of
      a property. The case is similar to the previous case.
    \end{description}

    \noindent
    We continue by examining each case of $\constrToConceptExpr$ individually:
    \begin{description}
      \item[$\constrToConceptExpr(\top) = \top$] The constraint evaluates to
        true for all nodes. If $I$ is a model, then {$\top^I
          \equiv {\toAtomicConcept[s]}^I$} is true.
        Due to the construction of $\constructed{\sigma}{I}$, $s$ is assigned to all nodes. As
        such, the shape is assigned to all nodes for which the constraint
        evaluates to true.

    \item[$\constrToConceptExpr(s') = \toAtomicConcept(s')$] The constraint evaluates to true
      for all nodes $o$ for which $s' \in \constructed{\sigma}{I}(o)$. Since $I$ is a model,
      $\toAtomicConcept(s')^I \equiv \toAtomicConcept[s]^I$ must be true. Due to the construction of
      $\constructed{\sigma}{I}$, $s$ is assigned to all nodes which are also assigned to $s'$.
      As such, the shape is assigned to all nodes for which the constraint
      evaluates to true.

    \item[$\constrToConceptExpr(v) = \{\toObjectName(v)\}$] The constraint evaluates only for
      the node $v$ to true. Since $I$ is a model, {$\{\toObjectName(v)\}^I \equiv
      {\toAtomicConcept[s]}^I$} must
      be true. Due to the construction of $\constructed{\sigma}{I}$, $s$ is only assigned to
      the node $v$.

    \item[$\constrToConceptExpr(\phi_1 \land \phi_2)
      = \constrToConceptExpr(\phi_1) \sqcap \constrToConceptExpr(\phi_2)$] The
      constraint evaluates to true for all nodes for which $\phi_1$ and $\phi_2$
      evaluate to true. By hypothesis, {$\constrToConceptExpr(\phi_1)
      = C$} and
      {$\constrToConceptExpr(\phi_2) = D$} represent the set of nodes for which
      $\phi_1$ and $\phi_2$ evaluate to true, respectively. Furthermore,
      {${(C\sqcap D)}^I \equiv {\toAtomicConcept[s]}^I$} is true if $I$ is a model. Due to the
      construction of $\constructed{\sigma}{I}$, the shape $s$ is assigned to all nodes in
      {${(C\sqcap D)}^I$}. The shape is therefore assigned to all nodes for which
      the constraint evaluates to true.

    \item[$\constrToConceptExpr(\neg \phi_1) = \neg \constrToConceptExpr(\phi_1)$]
      The constraint evaluates to true for all nodes for which $\phi_1$ does not
      evaluate to true. By hypothesis, {$\constrToConceptExpr(\phi_1)
      = C$}
      represent the set of nodes for which $\phi_1$ evaluates to true.
      Furthermore, since $I$ is a model, {$(\neg C)^I
      = {\toAtomicConcept[s]}^I$} must be
      true. Due to the construction of $\constructed{\sigma}{I}$, the shape $s$ is assigned to
      all nodes for which $\phi_1$ evaluates to false. The shape is therefore
      assigned to all nodes for which the constraint evaluates to true.

    \item[$\constrToConceptExpr(\geqslant_n\!\pe.\phi_1) = \geq_n\!r.\constrToConceptExpr(\phi_1)$]
      The constraint evaluates to true for all nodes that have $n$ successors
      via the path expression $\pe$ for which the constraint $\phi_1$ evaluates to true.
      By hypothesis, {$\constrToConceptExpr(\phi_1) = C$} is the set of nodes for
      which $\phi_1$ evaluates to true. Futhermore, since $I$ is a model,
      {$(\geq_n\!r.C)^I = {\toAtomicConcept[s]}^I$}. Due to the construction of $\constructed{\sigma}{I}$, the
      shape $s$ is assigned to all nodes that have $n$ successors and that are
      in the interpretation of {$C^I$}. The shape is therefore assigned to all
      nodes for which the constraint evaluates to true. \qedhere

  \end{description}
\end{proof}
}{
  \begin{proof}[Proof (Sketch)]
    The two axioms that are generated by $\shapesToKB$ coincide with the two
    conditions for faithful assignments (c.\,f.
    Definitions~\ref{def:faithful_assignment} and \ref{th:shapesToKB}). This can
    be shown by induction over the translation rules.
  \end{proof}
}

\subsection{Deciding Shape Containment using Concept Subsumption}

Given the translation rules and semantic equivalence between finite models of a
description logic knowledge base and assignments for SHACL shapes, we can
leverage description logics for deciding shape containment. Assume a set of
shapes $S$ containing definitions for two shapes $s$ and $s'$. Those shapes are
represented by atomic concepts in the knowledge base $\constructed{K}{S}$. As
the following theorem proves, deciding whether the shape $s$ is contained in the
shape $s'$ is equivalent to deciding concept subsumption between $s$ and $s'$ in
$\constructed{K}{S}$ using finite model reasoning.
\begin{theorem}[Shape containment and concept subsumption]
  \label{th:containmend_and_subsumption}
  Let $S$ be a set of shapes and $\constructed{K}{S}$ the knowledge base
  constructed via $\shapesToKB(S)$. Let $\models_\mathrm{fin}$
  indicate that an axiom is true in all finite models. It holds that:
  \[
    \contained{s}{s'}
    \Leftrightarrow \constructed{K}{S} \models_\mathrm{fin} \toAtomicConcept[s] \sqsubseteq
    \toAtomicConcept[s']
  \]
\end{theorem}
\ifExtended{
\begin{proof} Intuitively, the two problems are equivalent because any
  counterexample for one side of the equivalence relation could always be
  translated into a counterexample for the other side.
  If {$\constructed{K}{S} \not\models_\mathrm{fin} \toAtomicConcept[s] \sqsubseteq
  \toAtomicConcept[{s'}]$}, then there is a finite model of $\constructed{K}{S}$ in which
  {$\toAtomicConcept[s] \sqsubseteq \toAtomicConcept[{s'}]$} is not true
  Instead, there must be a model in which the concept expression
  {$\toAtomicConcept[s] \sqcap \neg \toAtomicConcept[{s'}]$} is true.
  Using Definition~\ref{def:containment:construction_assignment}, this model can
  be translated into an RDF graph and an assignment
  (c.\,f.~Theorem~\ref{th:constructed_assignment}) that acts as a counterexample
  for $s$ being contained in $s'$.  If {$\constructed{K}{S} \models_\mathrm{fin}
  \toAtomicConcept[s] \sqsubseteq \toAtomicConcept[{s'}]$}, then there is no
  finite model in which {$\toAtomicConcept[s]
  \sqsubseteq\toAtomicConcept[{s'}]$} is not true.
  Subsequently, there cannot be an RDF graph and an assignment that acts as
  a counterexample to $s$ being contained in $s'$, because this could be
  translated into a finite model of $\constructed{K}{S}$ using
  Definition~\ref{def:containment:constructed_interpretation}
  (c.\,f.~Theorem~\ref{th:constructed_interpretation}) and we know that no such
  model exists. \qedhere
\end{proof}
}{
  \begin{proof}[Proof (Sketch)]
    Using Theorems~\ref{th:constructed_interpretation} and
    \ref{th:constructed_assignment}, any counterexample for one side can
    always be translated to a counterexample for the other side. \qedhere
  \end{proof}
}
\noindent
As an example, reconsider the translation of the set of shapes
$\shapesToKB(S_1) = \constructed{K}{S_1}$ (see Fig.~\ref{fig:translation:s1}).
%
From $\constructed{K}{S_1}$ follows that
$\constructed{K}{S_1}\not\models\term{CubistShape}\sqsubseteq\term{PainterShape}$
as there is a finite model $I_1 \in \finmo{\constructed{K}{S_1}}$ in which the concept expression
$\term{CubistShape}\,\sqcap\,\neg\term{PainterShape}$ is
satisfiable (see Fig.~\ref{fig:interpretation_i1}).

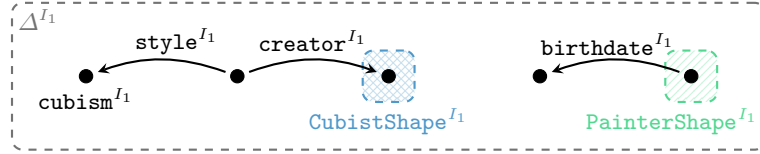
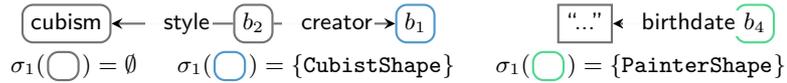
\begin{figure}[htp]
  \begin{subfigure}{1.0\textwidth}
    \setlength{\vgap}{0.75cm}
    \setlength{\hgap}{1.8cm}
    \centering
    \begin{tikzpicture}
      \node[object] (c1) {};
      \node[object,left=\hgap of c1] (p1) {}
      edge[roleout, bend left=20] node[above,midway]{${\footnotesize\term{creator}^{\tiny I_1}}$} (c1);
      \node[object,left=\hgap of p1] (s1) {}
      edge[rolein, bend left=20] node[above,midway,xshift=0.2cm, yshift=-0.05cm]{${\footnotesize\term{style}^{\tiny I_1}}$} (p1);

      \node[object, right=\hgap of c1] (bdate) {};
      \node[object, right=\hgap of bdate] (p2) {}
      edge[roleout, bend right=20] node[above,midway,xshift=-0.1cm, yshift=-0.05cm]{${\footnotesize\term{birthdate}^{\tiny I_1}}$}(bdate);

      \begin{scope}[on background layer]
        \node[below=0.05cm of s1, xshift=-0.0cm, yshift=0.1cm] {$\footnotesize\term{cubism}^{\tiny I_1}$};
        \node[text=darkgray] at (-4.6,0.75) {$\Delta^{I_1}$};

        \node[block, pattern color=highlight2!30, fit=(c1), draw=highlight2] (CubistShape) {};
        \node[block2, pattern color=highlight1!30, fit=(p2), draw=highlight1] (PainterShape) {};

        \node[below=0.05cm of CubistShape, xshift=-0.0cm, yshift=0.1cm, text=highlight2] {$\footnotesize\term{CubistShape}^{\tiny I_1}$};
        \node[below=0.05cm of PainterShape, xshift=-0.25cm, yshift=0.1cm, text=highlight1] {$\footnotesize\term{PainterShape}^{\tiny I_1}$};

        \node[oval, fit=(c1) (p1) (s1) (bdate) (p2)] (Universe) {};
      \end{scope}
    \end{tikzpicture}
    \subcaption{Model of $\constructed{K}{S_1}$ showing that $\term{CubistShape}\not\sqsubseteq\term{PainterShape}$.}
  \end{subfigure}

  \vspace{0.5cm}

  \begin{subfigure}{1.0\textwidth}
    \setlength{\vgap}{0.75cm}
    \setlength{\hgap}{1.6cm}
    \centering
    \begin{tikzpicture}
      \node[iri, draw=highlight2] (c1) {$b_1$};

      \node[iri, left=\hgap of c1] (p1) {$b_2$}
      edge[arrout] node[label]{creator} (c1);

      \node[iri, left=\hgap of p1] (s1) {cubism}
      edge[arrin] node[label]{style} (p1);


      \node[literal, right=\hgap of c1] (bdate) {``...''};
      \node[iri, right=\hgap of bdate, draw=highlight1] (p2) {$b_4$}
      edge[arrout] node[label]{birthdate} (bdate);
    \end{tikzpicture}

    \begin{tikzpicture}

      \node[] (sigmaempty) {\small $\sigma_1($};
      \node[smalliri,right=0.01em of sigmaempty,xshift=-0.1cm] {};
      \node[right=0.25cm of sigmaempty] {{\small $) = \emptyset$}};

      \node[right=1.5cm of sigmaempty] (sigmaCubist) {\small $\sigma_1($};
      \node[smalliri,right=0.01em of sigmaCubist,xshift=-0.1cm,draw=highlight2] {};
      \node[right=0.25cm of sigmaCubist] {{\small $) = \{\shapename{CubistShape}\}$}};

      \node[right=3.5cm of sigmaCubist] (sigmaPainter) {\small $\sigma_1($};
      \node[smalliri,right=0.01em of sigmaPainter,xshift=-0.1cm,draw=highlight1] {};
      \node[right=0.25cm of sigmaPainter] {{\small $) = \{\shapename{PainterShape}\}$}};



%
    \end{tikzpicture}
    \subcaption{Graph and assignment showing that
      \shapename{\small CubistShape} is not contained in \shapename{\small PainterShape}.}
  \end{subfigure}

  \caption{Counterexamples for $\shapename{CubistShape}~{<:}_{S_1}~\shapename{PainterShape}$.}
  \label{fig:interpretation_i1}
\end{figure}

An important observation is that it is possible to express arbitrary concept subsumptions $C
\sqsubseteq D$ despite the syntactic restrictions of $\shapesToKB$.
\begin{lemma}\label{lemma:subsumption_axioms}
  For any axiom $C \sqsubseteq D$, one can define some $(s,\phi,q) \in S$ and
  $(s',\phi',q') \in S$ such that $\constrToConceptExpr(\phi) = C$ and
  $\constrToConceptExpr(\phi') = D$ and $\shapesToKB(S) \models C \sqsubseteq
  D$.
\end{lemma}

\ifExtended{
\begin{proof}
  The concepts $C$ and $D$ used in the concept subsumption axiom $C \sqsubseteq
  D$ follow no syntactic restrictions, but that can use the complete syntax
  defined in Fig.~\ref{fig:dl:concept_expr}. The function $\shapesToKB$ on the
  other hand generates axioms of the following form:
  \[
    C^q \sqsubseteq A^\text{shape}, D^\phi \equiv A^\text{shape}
  \]
  where $A^\text{shape}$ is a atomic concept that represents a shape name,
  $D^\phi$ is a concept expression and $C^q$ is the translation of
  the target node query which adheres to the following grammar:
  \[
    C^q ::= \bot \mid \{v_1, \ldots, v_n\} \mid A^\text{class} \mid \exists\,p.\top \mid
    \exists\,p^-.\top
  \]
  As constraints $\phi$ use the same syntactical connectors as concept
  expressions (c.\,f. Section 2 and Fig.~\ref{fig:dl:concept_expr}), there is a
  $\phi_C$ such that $\constrToConceptExpr(\phi_C) = C$ and there is some $\phi_D$
  such that $\constrToConceptExpr(\phi_D) = D$. Furthermore, for both $C$ and $D$,
  we introduce shape names $A^\text{shape}_C$ and $A^\text{shape}_D$. Thus,
  $\shapesToKB$ will produce the axioms $C \equiv A^\text{shape}_C$ and $D \equiv
  A^\text{shape}_D$. To represent the inclusion, target nodes must be used. However, a
  shape such as $A^\text{shape}_C$ cannot act as a target node. Instead, we
  introduce an atomic concept $A^\text{class}_C$ that represents an RDF class
  and modify the constraint for shape $A^\text{shape}_C$ such that it includes
  the RDF class, giving us $C \sqcap A^\text{class}_C \equiv A^\text{shape}_C$.
  The shape $A^\text{shape}_D$ can then target $A^\text{class}_C$, completing
  the subsumption.

  In summary, the axiom $C \sqsubseteq D$ is semantically equivalent to the
  following set of axioms:
  \begin{align*}
    \{ C \sqcap {A^\text{class}_C} \equiv {A^\text{shape}_C}, &~\bot \sqsubseteq {A^\text{shape}_C} \\
       D \equiv {A^\text{shape}_D}, &~{A^\text{class}_C} \sqsubseteq {A^\text{shape}_D} \}
  \end{align*}
  It is therefore possible to represent any concept subsumption $C \sqsubseteq
  D$ through a set of shapes. \qedhere
\end{proof}
}{
  \begin{proof}[Proof (Sketch)]
    Given constraints $\phi$ and $\phi'$, it is possible to introduce unique
    shape names $s_C$ and $s_D$ as well as an RDF class $v_C$. Constraint $\phi$
    is then extended with $v_C$, allowing shape $s_D$ to target $v_C$.
  \end{proof}
}
For shapes belonging to the language $\mathcal{L}$, the corresponding
description logic is $\mathcal{ALCOIQ}(\circ)$. To the best of our knowledge,
finite satisfiability has not yet been investigated for
$\mathcal{ALCOIQ}(\circ)$.
Path concatenation can be restricted such that the fragment of SHACL corresponds
to the description logic $\mathcal{SROIQ}$. The fragment for which constraints
map to syntactical elements of $\mathcal{SROIQ}$, called
$\mathcal{L}^\restr$, uses the following constraint grammar:
\begin{align*}
  \phi^{\restr} ::=~& \top
                      \mid s
                      \mid v
                      \mid {\phi_1}^\restr \land {\phi_2}^\restr
                      \mid \neg {\phi}^\restr
                      \mid \exists\,\pe.\phi^\restr
                      \mid \geqslant_n\!{p}.{\phi}^\restr
\end{align*}
Finite satisfiability is known to be decidable for
$\mathcal{SROIQ}$~\cite{sroiqFinite} and all its sublogics such as
$\mathcal{ALCOIQ}$ which completely removes role concatenation.

\end{document}

\section{Deciding Shape Containment using Standard Entailment}
\label{sec:deciding_containment}

While shape containment can be decided using finite model reasoning
(c.\,f.~Theorem~\ref{th:containmend_and_subsumption}), practical usability of
our approach depends on whether existing reasoner implementations can be
leveraged. Implementations that are readily-available rely on standard
entailment which includes infinitely large models. We therefore now focus on the
soundness and completeness of our approach using the standard entailment
relation.   

Using standard entailment, the description logic $\mathcal{ALCOIQ}(\circ)$ which
corresponds to the language $\mathcal{L}$, satisfiability of concepts, and thus
concept subsumption, is undecidable~\cite{dlWithComposition}. 
First-order logic is semi-decidable. As $\mathcal{ALCOIQ}(\circ)$ can be
translated to first-order logic through a straightforward extension of the
translation rules for $\mathcal{SROIQ}$~\cite{foundationsDL}, $\mathcal{ALCOIQ}(\circ)$ is
also semi-decidable. 
Therefore, a decision procedure can verify whether a formula is entailed in
finite time, but may not terminate for non-entailed formula. 
More restricted description logics such as
$\mathcal{SROIQ}$, which corresponds to $\mathcal{L}^\restr$, are decidable,
meaning that an answer by the decision procedure is guaranteed in finite time.
However, the question arises whether the satisfiability of a concept implies the
existence of a finite model. 

%
%

%
\begin{definition}[Finite Model Property]
  A description logic has the {\normalfont finite model property} if every
  concept that is satisfiable with respect to a knowledge base has a finite
  model~\cite{introdl}.
\end{definition}
\noindent
If $C$ is a concept expression that is satisfiable with respect to some
knowledge base $K$ that belongs to a description logic having the finite model
property, then there must be a finite model of $K$ that shows the satisfiability
of $C$. Thus, finite entailment and standard entailment are the same if
a description logic has the finite model property. 

\begin{proposition}\label{prop:finiteModel:alcoiq}
  The finite model property does not hold for the description logic
  $\mathcal{ALCOIQ}$~\cite{finiteModel} or more expressive description logics
  such as $\mathcal{ALCOIQ}(\circ)$ and $\mathcal{SROIQ}$. If
  a concept expression $C$ is satisfiable with respect to a knowledge base $K$
  written in $\mathcal{ALCOIQ}$ or a more expressive description logic, then it may be that there are only models with
  an infinitely large universe.
\end{proposition}
\ifExtended{
To highlight Proposition~\ref{prop:finiteModel:alcoiq}, consider the following
example adapted from~\cite{finiteModel}:
\begin{alignat*}{5}
  K_\text{infinite} = \{ & \,\term{Painting}\, &\,\equiv\,&
  \,\exists\,\term{influences}.\term{Painting}~\sqcap& \\
  & & &
    \hspace{0.5cm}\leq\!1\,\term{influences}^-.\top,\, & \\
                         & \,\term{NovelPainting}\,&\,\equiv\,&\,\term{Painting}~\sqcap~
\leq\!0\,\term{influences}^-.\top \, & \}
\end{alignat*}
Each {\term{Painting}} influences another \term{Painting}, but is influenced by
at most one other \term{Painting}. A novel painting is a \term{Painting}
that is not influenced by anything. The concept {\term{NovelPainting}} is
satisfiable, but not finitely satisfiable. An object that is an instance of
{\term{NovelPainting}} must have a second object which it influences. This
second object must be an instance of {\term{Painting}} which means that it must
influence another instance of \term{Painting}. This leads to an infinite
sequence of paintings.
}{}

Given Proposition~\ref{prop:finiteModel:alcoiq}, it may be that
there are only models with an infinitely large universe that show the
satisfiability of a concept expression. There are three different possibilities:
(1) $\toAtomicConcept[s] \sqcap \neg \toAtomicConcept[{s'}]$ is neither finitely
nor infinitely satisfiable, meaning that
$\constructed{K}{S}\models\toAtomicConcept[s]\sqsubseteq\toAtomicConcept[{s'}]$.
It follows that $\contained{s}{s'}$ is true, as there is no counterexample. (2)
$\toAtomicConcept[s] \sqcap \neg \toAtomicConcept[{s'}]$ is not finitely, but
only infinitely satisfiable. It follows that
$\constructed{K}{S}\not\models\toAtomicConcept[s]\sqsubseteq\toAtomicConcept[{s'}]$,
but $\contained{s}{s'}$ is true since the infinitely large model has no
corresponding RDF graph. (3) $\toAtomicConcept[s] \sqcap \neg
\toAtomicConcept[s']$ is both, finitely and infinitely, satisfiable. It follows
that $\constructed{K}{S} \not\models
\toAtomicConcept[s]\sqsubseteq\toAtomicConcept[{s'}]$ and indeed
$\contained{s}{s'}$ is false since the finite model can be translated into an
RDF graph and a faithful assignment. Deciding shape containment for the shape
languages that are translatable into $\mathcal{ALCOIQ}(\circ)$, $\mathcal{SROIQ}$ or $\mathcal{ALCOIQ}$ is
therefore sound, provided that the decision procedure terminates.
\begin{theorem}
  Let $S$ be a set of shapes of the language $\mathcal{L}^\restr$.
  It then holds that:
  \begin{align*}
    \contained{s}{s'}
    \Leftarrow \shapesToKB(S) \models \toAtomicConcept[s] \sqsubseteq
    \toAtomicConcept[s']
  \end{align*}
\end{theorem}

\begin{proof}
  For $\mathcal{L}^\restr$, the corrseponding DL is $\mathcal{SROIQ}$ for which
  the finite model property does not hold. 
  If {$\constructed{K}{S} \models \toAtomicConcept[s] \sqsubseteq
    \toAtomicConcept[{s'}]$}, then there is neither a finitely nor an infinitely
  large model in which $\toAtomicConcept[s] \sqcap \neg\toAtomicConcept[s']$ is
  satisfiable.
  The shape $s$ must therefore be contained in the shape $s'$ as there is
  no RDF graph and assignment that acts as a counterexample.\qedhere
  %
\end{proof}
%
\noindent
However, the approach is incomplete as it may be that $\contained{s}{s'}$ but
$\constructed{K}{S}\not\models\toAtomicConcept[s]\sqsubseteq\toAtomicConcept[s']$
because due to an infinitely large model in which
$\toAtomicConcept[s]\sqcap\neg\toAtomicConcept[s']$ is satisfiable. 

To restore the finite model property, inverse path expressions have to be
removed. That is, the set of SHACL shapes $S$ must belong to the language
fragment $\mathcal{L}^\ninv$ that uses the following grammar:
\begin{align*}
  \phi^\ninv ::=~& \top
                   \mid v
                   \mid s
                   \mid {\phi_1}^\ninv \land {\phi_2}^\ninv
                   \mid \neg \phi^\ninv
                   \mid \geqslant_n\! p.\phi^\ninv \\
  q^\ninv ::=~& \bot \mid \{v_1, \ldots, v_n\} \mid \term{class}~v \mid \term{subjectsOf}~p
\end{align*}
As a result, the description logic that corresponds to $\mathcal{L}^\ninv$ is
$\mathcal{ALCOQ}$.
\begin{proposition}\label{prop:finiteModel:alcoq}
  The description logic $\mathcal{ALCOQ}$ has the finite model
  property~\cite{concreteDomains}.
\end{proposition}

Subsequently, for SHACL shapes that belong to $\mathcal{L}^\ninv$ shape
containment and concept subsumption in the knowledge base constructed from the
set of shapes are equivalent.
\begin{theorem}
  Let $S$ be a set of shapes belonging to $\mathcal{L}^\ninv$.
  Let $\constructed{K}{S}$ be the knowledge base constructed through $\shapesToKB(S)$.
  Then it holds that
  \begin{align*}
    \contained{s}{s'}
    \Leftrightarrow \constructed{K}{S} \models \toAtomicConcept[s] \sqsubseteq
    \toAtomicConcept[{s'}]
  \end{align*}
\end{theorem}

\ifExtended{
\begin{proof}
  If there is an RDF data graph and an assignment that acts as a counterexample
  for $s$ being contained in $s'$, then it can be translated into a finite model
  that shows that  $\constructed{K}{S} \not\models \toAtomicConcept[s] \sqsubseteq
  \toAtomicConcept[{s'}]$ (c.\,f. Theorem~\ref{th:containmend_and_subsumption}).
  On the other hand, there may be a model that acts as a counterexample showing
  that {$\constructed{K}{S} \not\models \toAtomicConcept[s] \sqsubseteq
  \toAtomicConcept[{s'}]$}. Since $\mathcal{ALCOQ}$ has the finite model
  property (c.\,f.~Proposition~\ref{prop:finiteModel:alcoq}), there must be
  a finite model that can be used as a counterexample. Therefore, a model
  exists that can be translated into an RDF data graph and an assignment such
  that $s$ is not subsumed by $s'$. \qedhere
\end{proof}
}{
  \begin{proof}[Proof (Sketch)]
    Due to $\mathcal{ALCOQ}$ having the finite model property, it is always
    possible to construct counterexamples for either side
    (c.\,f.~Theorem~\ref{th:containmend_and_subsumption}).
  \end{proof}
}

In summary, using standard entailment our approach is sound and complete for the
fragment of SHACL not using path concatenation or inverse path expressions. If
inverse path expressions are used, then the approach is still sound although
completeness is lost. Once arbitrary path concatenation is added, the resulting
DL becomes semi-decidable. While an answer is not guaranteed in finite time,
shape containment is still sound. 

\end{document}

\section{Related Work}
\label{sec:related_work}

Several constraint-based schema languages for RDF have been proposed before
SHACL. Among those are \cite{constraints1,constraints2}. To the best of our
knowledge, containment has not been investigated for those languages.
Additionally, SPIN\footnote{\url{http://spinrdf.org/}} proposed the usage of
SPARQL queries as constraints. When queries are used to express constraints, the
containment problem for constraints is equivalent to query containment.
ShEx~\cite{shexSemantics} is a constraint language for RDF that is inspired by
XML schema languages. While SHACL and ShEx are similar approaches, the semantics
of the latter is rooted in regular bag expressions. Validation of an RDF graph
with ShEx therefore constructs a single assignment whereas the SHACL semantics
used in this papers deals with multiple possible assignments. The containment
problem of ShEx shapes has been investigated in~\cite{shexContainment}. Due to
the specific definition of recursion in ShEx, any graph that conforms to the ShEx 
shapes will also conform to an equivalent SHACL definition. However, not all 
graphs that conform to SHACL shapes conform to equivalent ShEx shapes. 
It may be that a shape is contained in another shape in ShEx, but not 
in SHACL as there is a graph that can act as a counter-example for SHACL that 
does not conform to the ShEx shapes.

Similar to dedicated constraint languages, there have been proposals for the
extension of description logics with constraints. While standard description
logics adopts an open-world assumption not suited for data validation,
extensions inlcude special constraint axioms~\cite{icOWL1,icOWL2}, epistemic
operators~\cite{epistemicDL}, and closed predicates~\cite{closedPredicates}.
Constraints constitute T-Box axioms in these approaches, making constraint
subsumption a routine problem. 
%

Lastly, containment problems have been investigated for
queries~\cite{klug,containmentBagSemantics}. The query containment problem is
slightly different as result sets of queries are typically sets of tuples
whereas in SHACL we deal with conformance relative to faithful assignments.
Given an RDF graph and a set of shapes there may be several, different faithful
assignments. Operators available for SHACL are more expressive than operators
found in query languages for which subsumption has been investigated. In
particular, recursion is not part of most query languages.  There is
a non-recursive subset of SHACL that is known to be expressible as SPARQL
queries~\cite{shaclSPARQL}. When constraints are expressed as queries,
containment of SHACL shapes becomes equivalent to query containment. Recursive
fragments of SHACL, however, cannot be expressed as SPARQL queries.

%

\section{Summary}
\label{sec:summary}

In this paper, we have presented an approach for deciding SHACL shape containment by
translating the problem into a description logic subsumption problem. Our
translation allows for using efficient and well-known DL reasoning implementations
when deciding shape containment. Thus, shape containment can be used, for
example, in query optimization. 

We defined a syntactic translation of a set of shapes into
a description logic knowledge base. We then showed that finite models of this knowledge
base and faithful assignments of RDF graphs can be mapped onto each
other. 
Using finite model reasoning, this provides a sound and complete decision
procedure for deciding SHACL shape containment, although the decidability of
finite satisfiability in $\mathcal{ALCOIQ}(\circ)$ is still an open issue. As
part of future work, we plan to adapt the proof used by~\cite{sroiqFinite},
which comprises of a translation of $\mathcal{SROIQ}$ into a fragment of
first-order logic for which finite satisfiability is known.   
To ensure practical applicability, we also investigated the soundness and
completeness of our approach using standard entailment. 
Our findings are summarized in Fig.~\ref{fig:results}.

\begin{figure}[htpb]
\begin{center}
  \begin{tabular}{l c c c c }\toprule
    \textbf{SHACL Fragment} & \textbf{DL} &~\textit{Sound}~&~\textit{Complete}
                            & \textit{Terminates} \\ \midrule
    $\mathcal{L}$ & $\mathcal{ALCOIQ}(\circ)$ & Yes & No & Not guaranteed \\
    $\mathcal{L}^\restr$ & $\mathcal{SROIQ}$  & Yes & No & Yes\\
    $\mathcal{L}^\ninv$ & $\mathcal{ALCOQ}$   & Yes & Yes & Yes  \\
    \bottomrule
  \end{tabular}
\end{center}
  \caption{Soundness and completeness for deciding shape containment through
  description logics reasoning using standard entailment.}
  \label{fig:results}
\end{figure}

Our approach is sound and complete for the SHACL fragment $\mathcal{L}^\ninv$
that uses neither path concatenation nor inverse roles, as the finite model
property holds for the corresponding description logic $\mathcal{ALCOQ}$. Thus,
finite entailment and standard entailment are the same for this description
logic. The finite model property is lost as soon as inverse roles are added.
Using standard entailment, our procedure is still sound for the fragment
$\mathcal{L}^\restr$ which translates into $\mathcal{SROIQ}$ knowledge bases,
but is incomplete due to the possibility of a knowledge base having only
infinitely large models. Lastly, the SHACL fragment $\mathcal{L}$ translates
into $\mathcal{ALCOIQ}(\circ)$ knowledge bases. Our approach is sound, but
incomplete. However, due to the semi-decidability of the description logic, it
may be that the decision procedure does not terminate.

\subsubsection*{Acknowledgements.}
The authors gratefully acknowledge the financial support of project LISeQ
(LA 2672/1-1) by the German Research Foundation (DFG).

\section{Erratum and Correction}
\label{sec:summary}

The definition for faithful assignments
(Definition~\ref{def:faithful_assignment}) that has been published in this paper contains an error that has been pointed out
by~\cite{shaclFormalization}. The issue occurs in case of shapes where target
nodes are explicitly enumerated, but do not occur in the graph that is
validated. E.\,g., in case where $S$ looks as follows
\[
  S_2 = \{\shapename{MyShape},\geqslant_1\!\term{knows}.\term{charlie},\{\term{alice}\}\}
\]
and the data graph $G_1$ comprises only the triple \gedge{bob}{knows}{charlie}.
According to Definition~\ref{def:faithful_assignment}, $G_1$ is valid with
respect to $S_2$. However, using the translation rules in
Definition~\ref{def:targets2concepts}, the explicitly enumerated target nodes
are represented through the nominal concept $\{\term{alice}\}$, which is
a subset of the objects contained in the concept representing
\shapename{MyShape} (see Definition~\ref{th:shapesToKB}). Thus, there must be
faithful assignments for which no model of the description logic knowledge base
in our translation exists. 

The SHACL documentation\footnote{https://www.w3.org/TR/shacl/} does not clarify
whether explicitly enumerated target nodes missing in the data graph
constitutes an error. However, it does clarify that the evaluation of the query
for target nodes $\{\term{alice}\}$ over $G_1$ returns the node $\term{alice}$.
We believe that it is therefore reasonable to consider target nodes that are
explicitly enumerated but missing in the data graph as errors. The situation can
be solved by changing Definition~\ref{def:faithful_assignment} such that all
nodes returned by the evaluation of the query for target nodes must occur in the
data graph:

\begin{definition}[Faithful assignment]\label{def:faithful_assignment}
  An assignment $\sigma$ for a graph $G = (V_G, E_G)$ and a set of shapes $S$ is faithful,
  iff for each $(s, \phi, q) \in S$, it holds that:
  \begin{itemize}
    \item $s \in \sigma(v) \Leftrightarrow \evalc{\phi}$.
    \item $v \in \evalq{q} \Rightarrow s \in \sigma(v)$.
  \end{itemize}
\end{definition}

\bibliographystyle{splncs04}
\bibliography{references}

\end{document}